\documentclass[aip, pop, reprint, amsmath, amssymb]{revtex4-2}

\usepackage{amsmath}
\usepackage{siunitx}
\usepackage{graphicx}
\usepackage{bm}

\DeclareSIUnit\gauss{G}
\newcommand{\del}{\vec{\nabla}}
\newcommand{\texttilde}{\raise.17ex\hbox{$\scriptstyle\sim$}}

\draft 

\begin{document}
\renewcommand{\vec}[1]{{\bm{#1}}}

\preprint{PoP}

\title{Line-Tied Flux Rope Relaxation and Reconnection: A 3D Kinetic Case Study} 



\author{Joshua Pawlak}
\email[]{jpawlak@princeton.edu}
\affiliation{Princeton Plasma Physics Laboratory, Princeton, NJ 08540, USA}
\affiliation{Department of Astrophysical Sciences, Princeton University, Princeton, NJ 08544, USA\looseness=-1}

\author{James Juno}
\affiliation{Princeton Plasma Physics Laboratory, Princeton, NJ 08540, USA}

\author{Jason M. TenBarge}
\affiliation{Department of Astrophysical Sciences, Princeton University, Princeton, NJ 08544, USA\looseness=-1}

\date{\today}

\begin{abstract}
Magnetic flux ropes are ubiquitous magnetic structures found in plasmas ranging from astrophysical to laboratory. We employ a newly-developed parallel-kinetic-perpendicular-moment (PKPM) model to simulate the 3D interaction and evolution of two line-tied flux ropes at realistic laboratory plasma parameters, while retaining essential parallel kinetic physics in the system. We find that ropes undergo a current-dependent transition from a diamagnetic to paramagnetic regime, which we quantify with a simple analytic model. Although the macroscopic structural evolution qualitatively differs significantly between these regimes, analyzing the reconnection in proper field-aligned coordinates reveals that the underlying kinetic dynamics remain similar. Using the squashing factor and quasi-potential as diagnostics of 3D magnetic reconnection, we identify the formation of a quasi-separatrix layer and show that these quantities provide consistent metrics for reconnection rate and structure.

\end{abstract}

\pacs{}

\maketitle 

\section{Introduction}  
Magnetic flux ropes are coherent structures of braided magnetic field lines. \cite{priest1990} They are considered fundamental building blocks of large-scale magnetized plasma structures, ranging from the solar corona \cite{chen2017, liu2020} and planetary magnetospheres\cite{yoon2024,slavin2003} to laboratory plasma devices. \cite{gekelman2012, sun2010, myers2016} The dynamics of these structures are governed by complex interaction between MHD-scale forces and kinetic-scale dissipation. Flux rope interaction may facilitate magnetic reconnection, leading to rapid changes in magnetic topology and the conversion of magnetic energy into plasma kinetic energy. \cite{ji2022} Accordingly, understanding the details of these processes is essential to characterizing the overall energy balance in any magnetized plasma. 

Laboratory experiments such as those performed at the Large Plasma Device \cite{gekelman2016} (LAPD) at UCLA have successfully demonstrated and studied the formation and interaction of flux ropes in regimes where kinetic effects are significant.\cite{gekelman2012,dehaas2017,gekelman2016a,gekelman2018} However, the inherently three-dimensional structure of interacting flux ropes renders their study solely through experimental diagnostics challenging. Numerical simulation is therefore an essential contributor to understanding the evolution of these systems.

Simulating the dynamics of flux ropes in 3D presents a significant computational challenge. Fluid models can capture macroscopic evolution but fail to accurately capture the kinetic-scale physics governing the reconnection layer.\cite{stanier2015} Conversely, first principles-based approaches such as Particle-In-Cell (PIC) methods can accurately capture kinetic-scale reconnection physics\cite{daughton2014,stanier2015} but at considerable computational expense. In order to avoid numerical self-heating, the Debye length in these simulations must be fully resolved on the spatial grid.\cite{birdsall1991} This constraint becomes particularly problematic for experimental plasmas, where the large scale separation between the Debye length and particle inertial lengths makes simulations computationally intractable without artificially-modified parameters. Direct Vlasov equation solvers using discontinuous Galerkin methods\cite{juno2018, hakim2020,kormann2025} or spectral methods\cite{vencels2016,roytershteyn2018,koshkarov2021} avoid the numerical heating caused in PIC codes, but the required six-dimensional phase-space discretizations often similarly makes these simulations prohibitively expensive. 

To bridge this gap, we employ a newly-developed parallel-kinetic-perpendicular-moment (PKPM) model.\cite{juno2025} The PKPM model captures the essential kinetic physics of well-magnetized systems by solving a kinetic equation for the parallel distribution function while reducing the perpendicular evolution via a spectral expansion. This approach reduces the six-dimensional Vlasov equation to a set of four-dimensional equations, making it computationally feasible to apply to 3D systems. 

Even when 3D simulations can be performed, the correct procedure for quantitatively evaluating 3D reconnection is still debated. Unlike 2D reconnection where reconnection occurs across well-defined nulls or separatrices, reconnection in 3D may occur at multiple interacting sites across a layer which does not correspond to a magnetic seperatrix. These challenges have motivated the development of approaches which consider global magnetic topology, defining concepts like the quasi-seperatrix layer, \cite{priest1995} an extension of the 2D concept of separatrices which defines a region where field lines connectivity changes rapidly, and the quasi-potential, \cite{hesse2005} which considers contributions from the non-ideal electric field $E_\parallel$ across an entire field line to define a reconnection rate.

In this paper, we present 3D PKPM simulations of two interacting flux ropes with line-tied geometry from a non-equilibrium initial condition, with parameters based on the experiments performed on LAPD.\cite{gekelman2016a} The paper is organized as follows: in Sec. \ref{sec:model}, we introduce and sketch the derivation of the PKPM model and provide details on the initial and boundary conditions used. In Sec. \ref{sec:3d_diagnostics}, we discuss the 3D diagnostics employed to analyze our simulations. In Sec. \ref{sec:results}, we present and discuss results from simulations of flux ropes in two distinct regimes: a ``low-current" diamagnetic regime consistent with current experiments, and a ``high-current'' paramagnetic regime that explores higher current densities. We show that, despite important structural differences in these two cases, global analysis reveals that the physics which drives reconnection in the systems remains the same. Finally, we discuss and summarize the important results in Sec. \ref{sec:conclusions}.

\section{Model System} \label{sec:model}
\subsection{The PKPM Model}
The simulations presented here utilize a newly-developed parallel-kinetic-perpendicular-moment (PKPM) model. \cite{juno2025} This model solves a reduced Vlasov-Maxwell system by taking the species distribution function in the local fluid velocity frame and field-aligned coordinates, $f(\vec{x}, v_\parallel, v_\perp, \theta)$ and expanding in Fourier harmonics in gyroangle $\theta$ and a Laguerre series in $v_\perp$. 

Specifically, the model transforms the Vlasov equation for distribution function $f(\vec{x}, \vec{v}, t)$ into a frame moving with the bulk species fluid velocity $\vec{u}$, i.e.
\begin{align}
\begin{split}
    \bar{f}(\vec{x}, \vec{v}, t) \rightarrow \bar{f}(\vec{x}, \vec{v}', t),\\
    \vec{v}=\vec{v}' + \vec{u}(\vec{x},t),
\end{split}
\end{align}
and casting this distribution function (in the moving frame) in local field-aligned velocity coordinates,
\begin{equation}
    \bar{f}(\vec{x}, \vec{v}', t) = \bar{f}(\vec{x}, v_\parallel, v_\perp, \theta, t).
\end{equation}

We then expand this distribution function as a Fourier series in the velocity gyroangle $\theta$,
\begin{align} \label{eq:fourier}
\begin{split}
    &\bar{f}(\vec{x}, v_\parallel, v_\perp, \theta, t) = \bar{f}_0(\vec{x}, v_\parallel, v_\perp, t) \\
    &+ \sum_{n=1}^\infty \left[ \bar{f}_n^c(\vec{x}, v_\parallel, v_\perp, t)\cos n\theta) + \bar{f}_n^s(\vec{x}, v_\parallel, v_\perp, t)\sin n\theta) \right], 
\end{split}
\end{align}
and each of the terms in this Fourier series as a Laguerre series in the perpendicular velocity coordinate $v_\perp$, i.e. for $n=0$, 
\begin{equation}
    \bar{f}_0(\vec{x}, v_\parallel, v_\perp, t) = \sum_{l = 0}^\infty F_l(\vec{x}, v_\parallel, t) G_M(v_\perp, T_\perp)L_l\left( \frac{m v_\perp^2}{2T_\perp} \right),
\end{equation}
where $L_l(x)$ is the $l$-th Laguerre polynomial, and 
\begin{equation}
    G_M(v_\perp, T_\perp) = \frac{m}{2\pi T_\perp}\exp\left(-\frac{m v_\perp^2}{2T_\perp}\right). 
\end{equation}

The lowest-order PKPM system is defined as this expansion truncated at the $n=0$ Fourier harmonic and the $l=1$ Laguerre coefficient. In this case, we define 
\begin{equation}
    \mathcal{G} = \frac{T_\perp}{m}(F_0 - F_1),
\end{equation}
and the Vlasov equation reduces to the following two coupled four-dimensional kinetic equations for each species:

\begin{widetext}
    \begin{align}
        \frac{\partial F_0}{\partial t} &+ \del\cdot\left[(v_\parallel \hat{b} + \vec{u})F_0\right] + \frac{\partial}{\partial v_\parallel}\left[\hat{b}\cdot\left( \frac{1}{\rho}\del\cdot\vec{P} - v_\parallel\hat{b}\cdot\del\vec{u} \right)F_0 + \mathcal{G}\del\cdot\hat{b}\right] = 0 \label{eq:pkpmF} \\
        \begin{split} \label{eq:pkpmG}
        \frac{\partial \mathcal{G}}{\partial t} &+ \del\cdot\left[(v_\parallel \hat{b} + \vec{u})\mathcal{G}\right] + \frac{\partial}{\partial v_\parallel}\left[\hat{b}\cdot\left( \frac{1}{\rho}\del\cdot\vec{P} - v_\parallel\hat{b}\cdot\del\vec{u} \right) \mathcal{G} + \left[ 4\mathcal{G} - 2\frac{T_\perp}{m}F_0 \right]\frac{T_\perp}{m}\del\cdot\hat{b} \right] \\
        &= \left[ \hat{b} \cdot (\hat{b}\cdot\del\vec{u}) - \del\cdot(v_\parallel\hat{b} + \vec{u}) \right]\mathcal{G}
        \end{split}
    \end{align}
\end{widetext}

The system is closed via the fluid momentum equations, with the gyrotropic pressure tensor directly computed from the Laguerre coefficients:
\begin{align}
    &\frac{\partial\rho\vec{u}}{\partial t} + \del\cdot(\rho\vec{u}\vec{u} + \vec{P}) = \rho\frac{q}{m}[\vec{E} + \vec{u}\times\vec{B}], \label{eq:mom}\\
    &\vec{P} = p_\parallel\hat{b}\hat{b} + p_\perp(\vec{I} - \hat{b}\hat{b}),\\
    &p_\parallel = n T_\parallel = m\int v_\parallel^2F_0dv_\parallel,\\
    &p_\perp = n T_\perp = m\int \mathcal{G} dv_\parallel,
\end{align}
and with the electromagnetic field evolution from Maxwell's equations,
\begin{align}
    \frac{\partial \vec{B}}{\partial t} + \del\times\vec{E} &= 0,\\
    \epsilon_0\mu_0 \frac{\partial\vec{E}}{\partial t} - \del\times\vec{B} &= -\mu_0 \vec{J},\\
    \del\cdot\vec{E} &= \frac{1}{\epsilon_0}\sum_\sigma q_\sigma n_\sigma,\\
    \del\cdot\vec{B} &= 0,
\end{align}
coupled to the momentum equation through the plasma currents and densities,
\begin{align}
    \vec{J} &= \sum_\sigma\frac{q_\sigma}{m_\sigma}\rho_\sigma \vec{u}_\sigma,\\
    \rho_\sigma &= m_\sigma n_\sigma = m_\sigma \int F_{0_\sigma} dv_\parallel.
\end{align}

This model is implemented numerically in the \texttt{Gkeyll} simulation framework. Details on the implementation of numerical solvers for kinetic equations in \texttt{Gkeyll} can be found in Refs. ~\onlinecite{juno2018,hakim2020,juno2025}.

In the physical system studied here, electrons remain well-magnetized through nearly the entire reconnection layer due to the presence of a strong guide field.\textcolor{black}{\cite{egedal2002,le2009,swisdak2005,stanier2017} }The application of this reduced kinetic model is therefore well-founded and makes the problem computationally tractable.

The PKPM model also includes species self-collisions with a Dougherty-Fokker-Planck collision operator, \cite{lenard1958, dougherty1964} which in the PKPM system is written as
\begin{align}
    C[F_0]&=\nu\frac{\partial}{\partial v_\parallel}\left( v_\parallel F_0 + \frac{T}{m}\frac{\partial F_0}{\partial v_\parallel} \right),\\
    C[\mathcal{G}]&=2\nu\left( \frac{T}{m}F_0 - \mathcal{G} \right)+\nu\frac{\partial}{\partial v_\parallel}\left( v_\parallel \mathcal{G}+\frac{T}{m}\frac{\partial\mathcal{G}}{\partial v_\parallel} \right),
\end{align}
where $T = \frac{1}{3}(T_\parallel + 2T_\perp)$ and the collision frequency $\nu$ is taken as constant. \textcolor{black}{Note that this operator does not have velocity dependence in the collision frequency, and, in the current implementation, does not include inter-species collisions.}

\subsection{Initial Conditions}\label{sec:init_conds}
Previous theory and simulation work has shown that a non-equilibrium initial condition consisting of a collimated flowing electron population along a background magnetic guide field against a stationary background may coalesce into a flux rope. \cite{yoon2024} Accordingly, we begin by defining a background field $\vec{B}_0 = B_0\hat{z}$ and initialize two nonequilibrium rope structures via the following \textit{lab-frame} PKPM distribution functions:
\begin{align}
    F_{0_e}(\vec{x},v_\parallel) =& F_{0_{e,b}} + F_{0_{e,r}}, \label{eq:F0}\\ 
    \mathcal{G}_e(\vec{x},v_\parallel) =& (T_{e,b} F_{0_{e,b}} + T_{e,r}F_{0_{e,r}})/m_e, \label{eq:G}
\end{align}
\noindent where the distribution function components are both Maxwellian with different relative flows representing the background and rope, respectively:
\begin{align}
    &F_{0_{e,b}}(\vec{x},v_\parallel) = n_b(\vec{x})\sqrt{\frac{m_e}{2\pi T_{e,b}}}\exp\left(-\frac{m_e {v_\parallel}^2}{2T_{e,b}}\right), \label{eq:background} \\
    &F_{0_{e,r}}(\vec{x},v_\parallel) = n_r(\vec{x})\sqrt{\frac{m_e}{2\pi T_{e,r}}}\exp\left[-\frac{m_e}{2T_{e,r}}(v_\parallel - \vec{V}_{0_e}\cdot\hat{b})^2\right]. \label{eq:rope}
\end{align} 
Here, $n_b(\vec{x})$ and $n_r(\vec{x})$ designate the background and rope plasma density profiles, respectively, and $T_{e,b}$, $T_{e,r}$ designate the corresponding initially-uniform temperatures of these two populations (in both parallel and perpendicular directions). $\vec{V}_{0_e}$ is the lab-frame fluid velocity of the rope population. The ion species uses an identical initial condition but with different temperatures $T_{i,b}$ and $T_{i,r}$ and no initial drift fluid velocity (i.e. $\vec{V}_{0_i} = 0$). The LAPD experiments report nearly-Gaussian radial temperature, density, and current profiles of the flux ropes against a uniform background with a sharp cutoff near the device wall. \cite{gekelman2018} Accordingly, we choose the plasma density profiles,
\begin{align}
    n_b(\vec{x}) = &\frac{n_{b,0}}{2}\left[1 - \tanh{\left(\frac{r - L_r}{\rho_s}\right)}\right] + a n_{b,0},\\
    \begin{split} \label{eq:nr}
    n_r(\vec{x}) = &n_{r,0} \left[\exp\left(-{|\vec{x} - \vec{x}_1|}^2/{r_s}^2\right) \right.\\
    & + \left.\exp\left(-{|\vec{x} - \vec{x}_2|}^2/{r_s}^2\right)\right].
    \end{split}
\end{align}
Here, $n_{b,0}$ and $n_{r,0}$ are the peak background and rope plasma densities, $L_r$ is the radius of the background plasma column, $\rho_s = \sqrt{T_{e,b}/m_i}/\Omega_{ci}$ is the ion sound gyroradius, $\vec{x}_1$ and $\vec{x}_2$ are the positions of the two flux ropes, and $r_s$ is the characteristic flux rope radius. The $\tanh$ dependence in $n_b$ defines a ``top hat" radial profile which quickly vanishes outside of the radius $r=L_r$. $a n_{b,0}$ defines an additional small uniform density on top of the sharp gradient, with $a=0.1$, to avoid encountering numerical errors that arise at very small plasma densities near the boundary. We note that $L_r$ is chosen to be sufficiently large relative to the flux rope separation that the background profile does not affect the flux rope dynamics directly, and the small density in the outer region results in negligible effects from the boundaries on the relevant flux rope region. However, we found this ``top hat" density profile to be important in suppressing electrostatic modes excited by rapid electron motion at very early times in the simulation. Similar density profiles are observed in LAPD plasmas,\cite{gekelman2016} so implementing this profile is also physically reasonable.

We normalize the drift velocity of the rope population $\vec{V}_{0_e}$ by fixing the total machine discharge current $I_0$, which is split between the two ropes. Using Eqs. \ref{eq:F0} and \ref{eq:nr}, we take the current density,
\begin{equation} \label{eq:current}
\vec{J}(\vec{x}) = -e n_r(\vec{x})V_{0_e}\hat{z},
\end{equation}

\noindent and integrate to find the total current (split equally between the two ropes) as $I_0 = 2\pi {r_s}^2 n_r e V_{0_e}$, which determines the drift velocity. 

The electric field is initially zero everywhere, $\vec{E}=0$. To determine the initial condition for the magnetic field, we assume $\del\times\vec{B} = \mu_0\vec{J}$ initially and solve for $\vec{B}$ using Eqn \ref{eq:current}. For a single, centered flux rope carrying total current $I$, this field is
\begin{equation} \label{eq:magfield}
    \vec{B}(\vec{x}) = [1 - \exp(-r^2/r_s^2)]\frac{\mu_0 I}{2\pi r}\hat{\phi}.
\end{equation}

Finally, for computational purposes, we choose an artificial mass ratio of $m_i/m_e = 400$ (fixing $m_i=4m_p$, the mass of $^4$He), and an artificially-reduced speed of light by increasing the vacuum permittivity constant, $\epsilon_{sim} = 1000\epsilon_0$. 

To mitigate the propagation of non-physical cavity modes which are triggered by the rapid relaxation of the initial non-equilibrium plasma, we introduce a spatially-dependent artificial damping term to the electric field evolution. This damping is large only outside the primary background plasma column and near the $z=\qty{0}{\meter}$ boundary away from the flux rope centers. 

Since the PKPM model used here only includes the $n=0$ Fourier harmonic from the expansion shown in Eq. \ref{eq:fourier}, it cannot correctly capture non-gyrotropic motion which will occur in a reconnecting plasma. \textcolor{black}{Previous particle-in-cell simulations of flux rope merging found that the non-gyrotropic pressure components played an important role in supporting the reconnection electric field in a parameter regime where the scale separation between the electron gyroradius and inertial length was $\rho_e/d_e=0.3$.\cite{sauppe2017} However, our work here considers a much larger scale separation, $\rho_e/d_e\approx0.043$, identical to that in the LAPD experiment, \cite{gekelman2016a} in which the electron gyroradius spans a much smaller part of the reconnection layer and is significantly under-resolved in our spatial discretization. The gyrotropic components of the pressure tensor still provide the main anisotropy in the system, with non-gyrotropic components only manifesting in a truly grid-scale numerical response across the grid cells where $B_x$ and $B_y$ vanish in the reconnection layer.\cite{le2009}} To address this limitation, we introduce a small non-physical 4th order hyper-diffusive term to the species equation of motion (Eq. \ref{eq:mom}). This diffusion is formulated as a term $D_{hyp}\nabla^4(\rho \vec{u})$, where $D_{hyp} = \nu_{hyp}(\omega_{pe}/2\pi)d_e^4$ and $\nu_{hyp} = 10^{-3}$ as a dimensionless scale quantity. Previous reconnection simulations using PKPM showed that this approach correctly smooths over the nonphysical response that results from this limitation in the model, without modifying the larger-scale kinetic response and layer width.\cite{juno2025,le2009} We also note that the optimal value of $\nu_{hyp}$ we found here is consistent with the value studied in that work.

In Appendix \ref{sec:init_details}, we present a complete tabulation of all relevant simulation parameters and expressions for initial conditions, including realistic laboratory plasma temperatures and densities. The physical parameters are chosen to closely match the LAPD experiment. We also show how the lab-frame distribution functions given in Eqs. \ref{eq:F0} and \ref{eq:G} are transformed into the local plasma rest frame implemented explicitly in PKPM. 

We perform 3D-1V simulations on a spatial grid of $128\times128\times140$ cells that spans a physical domain of $\qty{0.8}{\meter}\times\qty{0.8}{\meter}\times\qty{10}{\meter}$ and a 16-cell discretization in $v_\parallel$ spanning a velocity space from $-6(T_{\sigma,b}/m_\sigma)^{1/2}$ to $6(T_{\sigma,b}/m_\sigma)^{1/2}$, defined using the background temperature of each species, $T_{\sigma,b}$. We use linear polynomials in configuration space and quadratic polynomials in velocity space identical to the original PKPM paper \cite{juno2025} and other recent DG models of magnetized plasmas. \cite{francisquez2026}

\subsection{Boundary Conditions}

\paragraph{Electromagnetic Fields.}
We implement perfectly-conducting wall boundary conditions for both the electric and magnetic fields at the $z=\qty{0}{\meter}$ plane and all transverse (side) walls. To simulate a free boundary at $z=\qty{10}{\meter}$, we utilize a ``copy" boundary condition. The copy condition populates ghost cells by duplicating values from the interior boundary grid cells, allowing for the calculation of fluxes at the boundaries within the DG scheme.

\paragraph{Plasma Species.}
For the ion and electron populations, we utilize the same copy boundary condition described above for the boundary $z=\qty{10}{\meter}$ and all side walls. While a reflecting wall might seem more physically intuitive for the side boundaries, we found it led to non-physical charge accumulation and convergence failures at the corners of the Cartesian domain. Given that these boundaries are geometrically distant from the flux ropes and reside in regions of very low plasma density, this choice does not significantly impact the core dynamics.

At the $z=\qty{0}{\meter}$ boundary, we implement a ``line-tied" condition by enforcing a zero parallel gradient and zero perpendicular flow for the momentum ($\rho_\sigma \vec{u}_\sigma$) of each species. Although numerical diffusion results in minimal movement at the flux rope footpoints over time, the primary physical interactions of interest occur well before this effect becomes relevant.

\section{3D Diagnostics} \label{sec:3d_diagnostics}

Analyzing the outputs of three-dimensional reconnection simulations is challenging due to the complex magnetic topology which can develop as magnetic field lines twist and reconnect. No single separatrix point exists in 3D, and identifying the reconnection layer structure along the third dimension becomes nontrivial. To assist in interpreting the results of our 3D simulations, we implemented two well-established diagnostics: the squashing factor and the quasi-potential.

\subsection{Squashing Factor}
Field lines in a strong guide field define a coordinate mapping from initial coordinates $(x,y)$ at one $z=z_0$ plane to their final coordinates $(X(x,y), Y(x,y))$ at a second $z=z_f$ plane. The squashing factor $Q$ is a measure of the stretch and squeeze of field lines.\cite{wendel2013} It is defined by the expression,
\begin{equation} \label{eq:squashing_factor}
    Q = \frac{\left(\frac{\partial X}{\partial x}\right)^2 + \left(\frac{\partial X}{\partial y}\right)^2 + \left(\frac{\partial Y}{\partial x}\right)^2 + \left(\frac{\partial Y}{\partial y}\right)^2}{|B_z(z)/B_z(Z)|}.
\end{equation}

Regions with  high values of $Q$ indicate field lines which converge and/or diverge rapidly from one another. Collecting all field lines with a given value of $Q$ defines a surface known as the quasi-separatrix layer (QSL), which indicates places where 3D reconnection events are likely to occur.\cite{priest1995}

To implement this diagnostic, we interpolate our simulated magnetic field output data and integrate field lines from a grid of points in the $z=z_0$ plane to the $z=z_f$ plane \textcolor{black}{(here, using an $8000\times8000$ grid of seed points with $z_0 = \qty{0}{\meter}$ and $z_f = \qty{10}{\meter}$)} using a symplectic second-order leapfrog method and calculate $Q(x,y)$ based on the map $(x,y)\rightarrow(X,Y)$ these field lines define. The QSL is then identified as the set of field lines where $Q$ exceeds some chosen threshold value. \textcolor{black}{Further details on this calculation and the interpretation of $Q$ are discussed in Appendix \ref{sec:squashing}.}

\subsection{Quasi-Potential}

Another useful diagnostic for three-dimensional reconnection is the quasi-potential $\Xi$.\cite{hesse1993,hesse2005} This quantity is defined as the integral of the electric field along a magnetic field line:
\begin{equation}\label{eq:quasipotential}
    \Xi = -\int_{z_0}^{z_f}\vec{E}\cdot \vec{dl}.
\end{equation}

In order to compute this quantity, we use the same field line-integrating scheme outlined above before using a trapezoid method to integrate the electric field along the computed field line paths. To avoid boundary effects, we begin at $z_0 = \qty{1}{\meter}$ and integrate to the end of the domain. We thus obtain a function $\Xi(x, y)$, where $(x,y)$ is the location of the field line at the $z=z_0$ plane.

The quasi-potential may be broken down further into the fluid terms which support it. Defining the bulk velocity $\vec{u}=(m_e\vec{u}_e + m_i\vec{u}_i)/(m_e + m_i)$, we refer to the generalized Ohm's law,
\begin{align} \label{eq:ohms_law}
    \begin{split}
        \vec{E} =& -\vec{u}\times\vec{B} - \frac{1}{ne}\del\cdot\vec{P}_e + \frac{1}{ne}\vec{J}\times\vec{B} \\
        &+\frac{m_e}{ne^2}\left[ \frac{\partial\vec{J}}{\partial t} + \del\cdot\left( \vec{u}\vec{J} + \vec{J}\vec{u} - \frac{\vec{J}\vec{J}}{ne} \right) \right].
    \end{split}
\end{align}
Noting the parallel component of $\del\cdot\vec{P}_e$ in the gyrotropic limit relevant to the PKPM model,
\begin{equation}
    \hat{b}\cdot(\del\cdot\vec{P}_e) = \hat{b}\cdot\del P_{e,\parallel} - (P_{e,\parallel} - P_{e,\perp})\hat{b}\cdot\frac{\del B}{B},
\end{equation}
we integrate both sides of Eq. \ref{eq:ohms_law} along a magnetic field line to obtain 
\begin{align} \label{eq:quasipotential_ohm}
    \begin{split}
    \Xi =& \int \frac{1}{ne}\left\{\del P_{e,\parallel} - (P_{e,\parallel} - P_{e,\perp})\frac{\del B}{B}\right.\\
    &\left.- \frac{m_e}{e}\left[\frac{\partial\vec{J}}{\partial t} - \del\cdot\left( \vec{u}\vec{J} + \vec{J}\vec{u} - \frac{\vec{J}\vec{J}}{ne} \right)\right]\right\}\cdot\vec{dl}.
    \end{split}
\end{align}

We thus see that the quasi-potential can only be supported by the parallel pressure tensor gradient, a pressure anisotropy term dependent on the gradient of the magnetic field strength, and inertial terms. Thus, this diagnostic is a useful way of identifying field lines that pass through the regions of non-ideal plasma motion which facilitate field line breaking. Previous theoretical work showed that the maximum quasi-potential is equal to the 3D reconnection rate. \cite{hesse2005} 

\section{Simulation Results} \label{sec:results}
\begin{figure*}[t]
\includegraphics[trim={5pt 5pt 5pt 5pt},clip,width=1\linewidth]{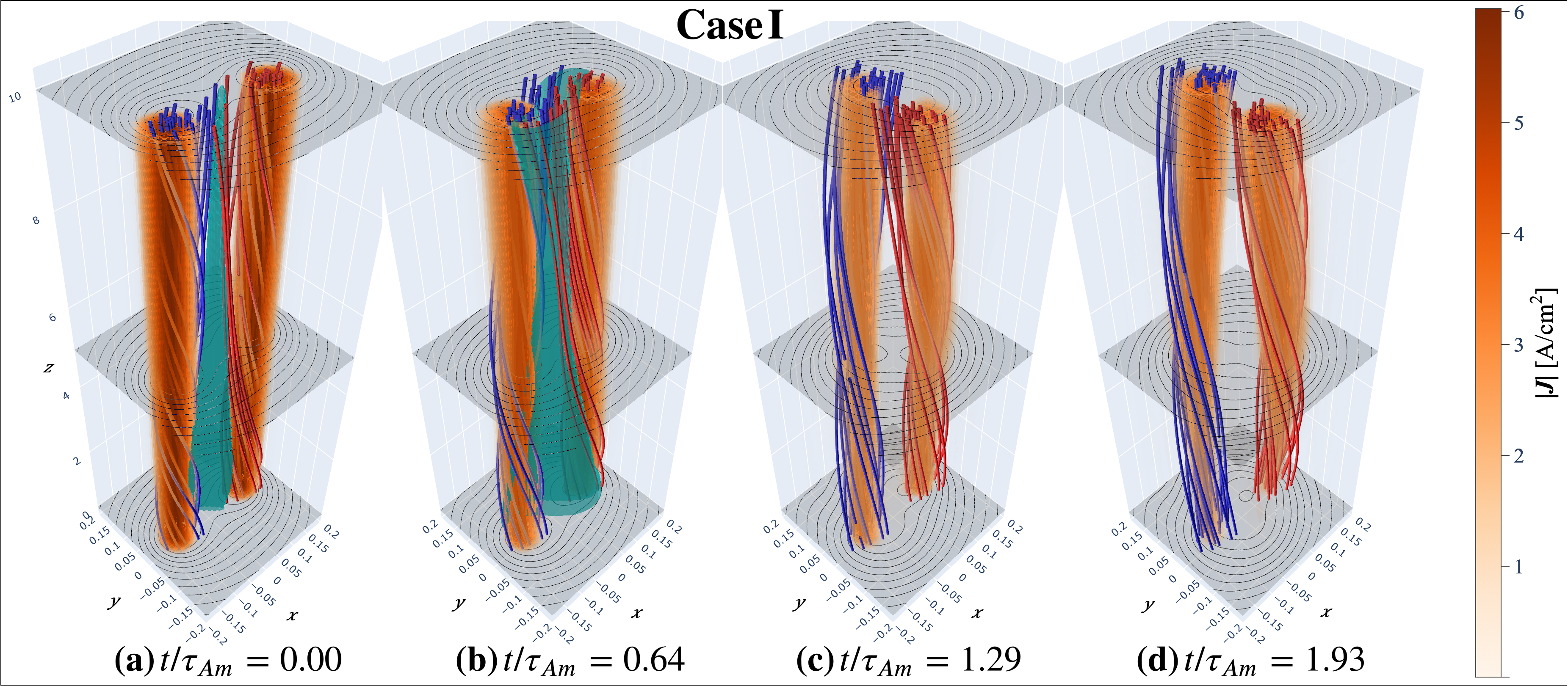}
\includegraphics[trim={5pt 5pt 5pt 5pt},clip,width=1\linewidth]{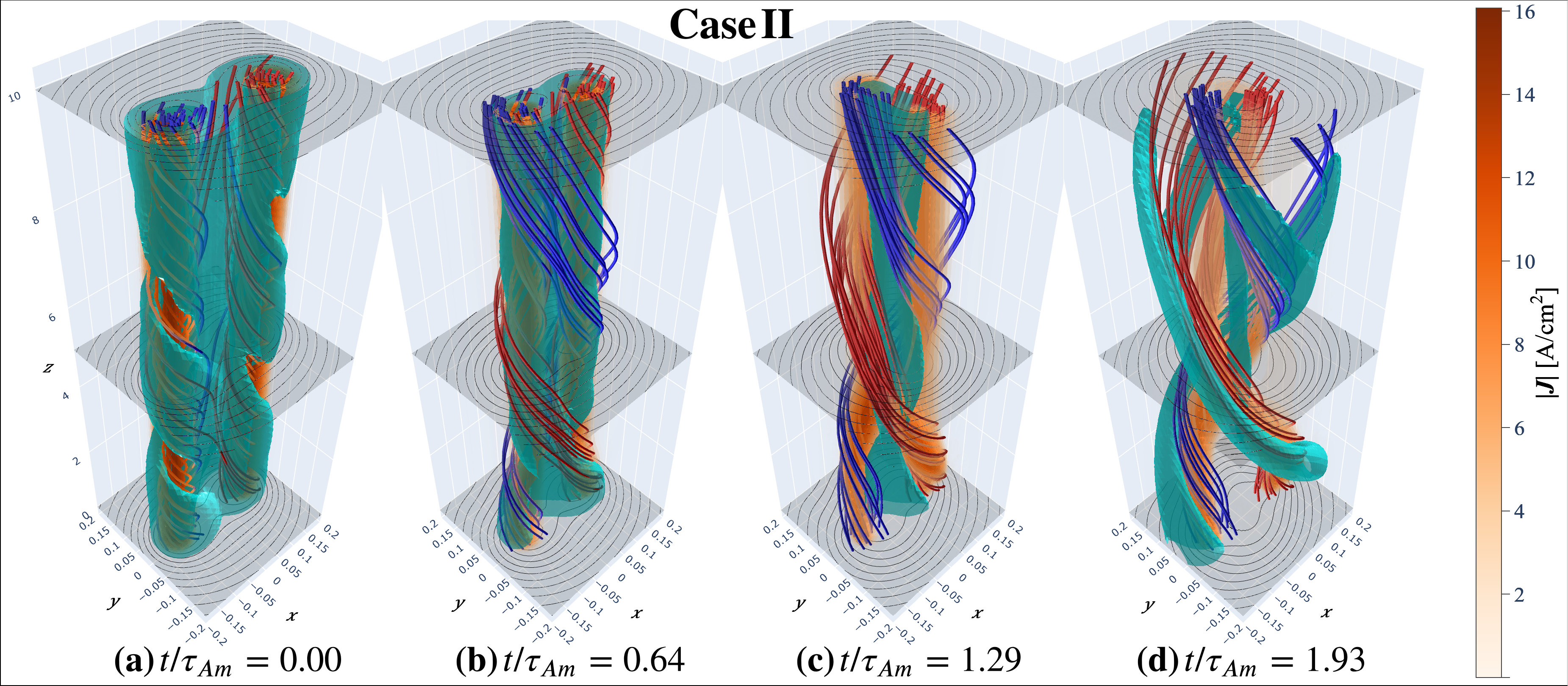}%
\caption{3D structure of the interacting flux ropes in the low-current case (case I, top) and high-current case (case II, bottom). Representative field lines are drawn originating at $z=\qty{0}{\meter}$ near the flux rope centers, with the blue or red color indicating which center the line started near. The volume-filling orange color indicates current magnitude in $\unit{\ampere\per\square\centi\meter}$, and the cyan surface indicates the QSL structure by a surface of constant squashing factor, $Q=1000$. Contours of the 2D flux function $\psi$ are also indicated on 2D cuts at $z=0$, $5$, and $\qty{10}{\meter}$.}
\label{fig:3d}
\end{figure*}

We ran two simulations of flux rope merging. One simulation was performed for a ``low-current" discharge case (``case I") with total current $I_0=\qty{-750}{\ampere}$ measured from the $z=\qty{0}{\meter}$ boundary, split between the two ropes; this case corresponds to the experiments performed on the LAPD platform. \cite{gekelman2016a} The second was performed for a ``high-current" discharge case (``case II") with total current $I_0=-\qty{2000}{\ampere}$. Case II is notable as it corresponds to a maximum current density of roughly $\qty{-16}{\ampere/\centi\meter^2}$, which is higher than the maximum current density $\qty{-10}{\ampere/\centi\meter^2}$ that can be achieved using the LaB$_6$ emissive cathode on LAPD. \cite{gekelman2016} This scenario allows case I to be directly benchmarked against existing experiments, and case II can thus be used to demonstrate speculative simulations into plasma parameters which have so far been understudied in laboratory experiments.

Due to the difference in current magnitude between the two simulations, the timescales of flux rope relaxation and their mutual interaction are different. We normalize the simulation times to the in-plane Alfv\'en crossing time, $\tau_{Am}\equiv L_{sep}/V_{Am}$, where $L_{sep}$ is the initial separation distance between the two flux ropes and $V_{Am}=B_m/\sqrt{m_i \overline{n}_{0} \mu_0}$ is defined with the mean plasma density $\overline{n}_0$ and maximum in-plane magnetic field $B_m$ from Eq. \ref{eq:magfield}. This choice is observed to properly normalize the timescale of in-plane motion observed in the simulation and is consistent with normalization choices used in previous work on guide field reconnection.\cite{stanier2015,stanier2017,sauppe2017}

Figure \ref{fig:3d} shows the general 3D structure of the flux ropes at several time snapshots as they evolve. Current magnitudes are indicated by the orange color, and several representative magnetic field lines are drawn originating near the initial rope centers at $z=\qty{0}{\meter}$. The red and blue color of the field lines indicates at which rope the field line began. 2D flux function contours are indicated along $x$-$y$ planes at $z=\qty{0}{\meter}$, $z=\qty{5}{\meter}$, and $z=\qty{10}{\meter}$. Note that while we use the 2D flux function, defined by $\vec{B}_\perp = \del{\psi}\times\hat{z}$, the strong guide field in our simulation leads to negligible differences from the full 3D definition. As the flux ropes merge and reconnect, a QSL forms, shown by the cyan surface. In both case I and case II, we observe the flux ropes to rapidly relax to their individual local equilibrium states before 3D structure develops as the two flux ropes rotate and partially merge with each other. 

{
\color{black}
The squashing factor $Q$ is, in general, highly peaked during reconnection, with values of $Q$ reaching $\texttilde 10^5$ in case I and $\texttilde 10^6$ in case II. Indicating a surface with this high of a threshold does not substantially change the overall structure, however, and makes visualization of its formation at early time more difficult. We choose a relatively low threshold value of $Q$ to illustrate this structure; discussion of the time dependence of the squashing factor is included in Sec. \ref{sec:recon_rate}.
}

The initially-straight flux ropes are attracted via the $\vec{J}\times\vec{B}$ force, to which electrons respond quickly. This reaction creates an in-plane current which interacts with the background guide field as an additional $\vec{J}\times\vec{B}$ component to drive the initial rotation of the ropes. Due to the line-tied boundary condition at $z=\qty{0}{\meter}$, this rotation cannot propagate across the entire domain, causing the observed helical structure to develop. 

\begin{figure*}[t]
\includegraphics[clip,width=1\linewidth]{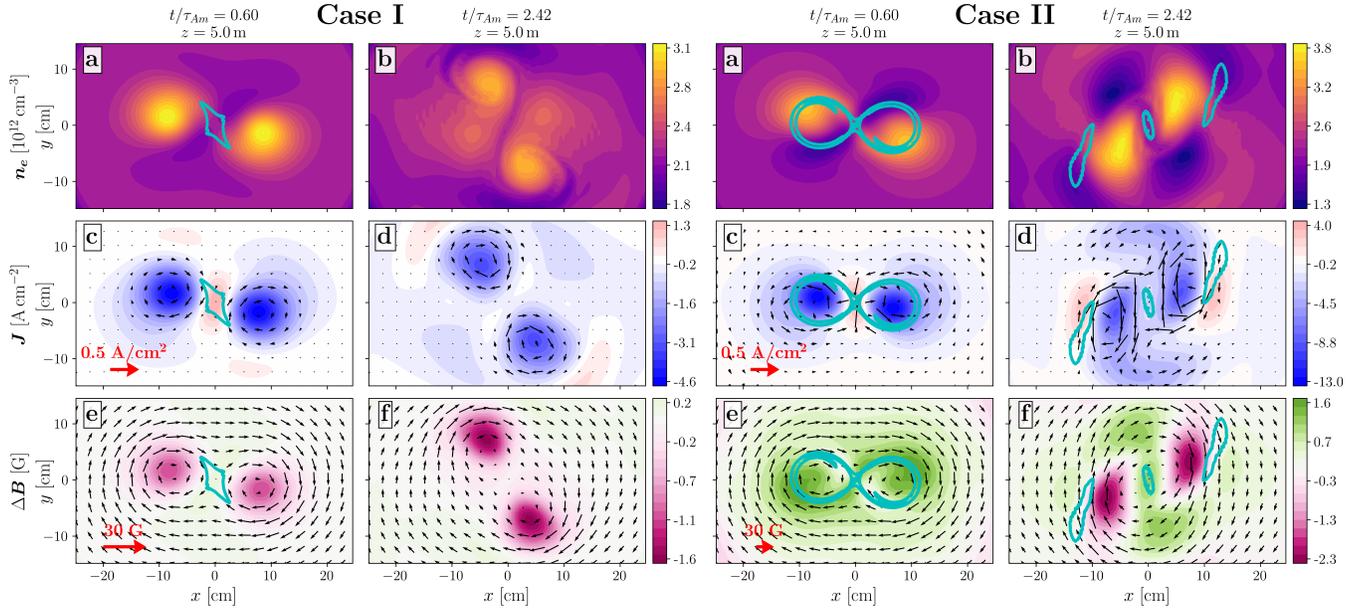}
\caption{2D cuts in the $x$-$y$ plane of the low-current (case I) and high-current (case II) 3D simulations at the $z=\qty{5}{\meter}$ plane at early time (a, c, e) and late time (b, d, f) showing plasma density (a, b), current density (c, d), and magnetic field deviation from the background, $\Delta\vec{B} = \vec{B} - B_0\hat{z}$ (e, f). For vector quantities $\vec{J}$ and $\vec{B}$, the out-of-plane ($\hat{z}$) component is indicated by the color, while the in-plane component direction and magnitudes are indicated by the arrow direction and lengths. The cyan contour indicates a contour of constant squashing factor $Q=500$ to show the QSL structure. Note that at late time, the squashing factor has dropped below this value everywhere in case I. }
\label{fig:2d_cuts}
\end{figure*}

To illustrate the internal structure and evolutionary dynamics of the interacting flux ropes, Figure \ref{fig:2d_cuts} presents the density, current, and magnetic field distributions for an $x$-$y$ plane slice at $z=\qty{5}{\meter}$. Snapshots are provided for both an early time, during active reconnection, and a late time, after reconnection has concluded. As the ropes merge, a current sheet develops in the opposite direction to the ropes, with current densities on the order of $\texttilde25$\% of the main flow (Figs. \ref{fig:2d_cuts}I.c, \ref{fig:2d_cuts}II.c). The cyan contour drawn in each subplot, indicating the QSL as defined by a surface of constant squashing factor $Q=500$, shows where field lines are most rapidly rearranging. Especially for case I, this region matches well to the current sheet structure at early time.

At late time, reconnection has largely completed, as indicated by the lack of a QSL between the two ropes in Fig. \ref{fig:3d}. At this time, the ropes begin to take on a helical structure. To understand this structure, we apply an analytic model for a MHD double helix equilibrium,\cite{zhang2025} which considers the balance of the attractive force between parallel axial current components of the two ropes with the repulsive force between anti-parallel in-plane current components of sufficiently-twisted ropes in a guide field. This calculation gives equilibrium wavelengths of $\lambda\approx\qty{67}{\meter}$ and $\lambda\approx\qty{25}{\meter}$ for the initial conditions in cases I and II, respectively. Though neither of our simulations demonstrate a complete helical rotation of the two ropes within the simulated domain to establish a rigorous wavelength, extrapolating from the current density structure within the domain corresponds to axial wavelengths of $\lambda\approx\qty{31}{\meter}$ in case I and $\lambda\approx\qty{10}{\meter}$ in case II. This result is somewhat shorter than the predicted equilibrium wavelengths, suggesting that the ropes have not yet reached an MHD equilibrium and that repulsive forces should be dominant; this is consistent with the dynamic evolution we observe at late time in our simulations.

Related simulation work on flux rope interaction has observed the growth of an ideal kink mode which plays a role in driving flux rope rotation and reconnection. \cite{sauppe2017} {\color{black} While the flux ropes in our simulations are kink-unstable, we do not observe the presence of a kink mode in either of our simulations. Theoretical work on the kink instability growth rate in flux ropes \cite{ryutov2006} shows that the growth time is larger than the merging timescale in our simulation, and the total simulation time is not sufficient to observe the characteristic oscillation frequency in LAPD experiments \cite{gekelman2016a}. Details on this timescale comparison and further simulation results demonstrating the growth of a kink instability in a single flux rope, may be found in Appendix \ref{sec:kink_modes}. We defer a careful analysis of kink modes in this two rope system, especially under more realistic experimental conditions with sheath boundary conditions and finite resistivity, to future work.
}

\subsection{Flux Rope Diamagnetism and Paramagnetism}

Case I, matching closely to LAPD experimental parameters, is observed to develop cross-field diamagnetic currents in the two flux ropes in the initial relaxation prior to merging, generating magnetic field perturbations at the centers of the ropes opposite to the guide field direction. These currents develop due to the pressure gradients within each flux rope. The peak diamagnetic current is approximately \qty{0.44}{\ampere/\centi\meter^2} and is consistent with measurements of diamagnetic current in the LAPD experiment. \cite{vancompernolle2012}

As seen in Fig. \ref{fig:2d_cuts}, a surprising transition occurs for case II, where the azimuthal current in the flux ropes reverses direction, leading to paramagnetic flux ropes after the initial relaxation. \textcolor{black}{This transition is related to well-known behavior in screw pinch equilibria.\cite{bickerton1958} The behavior has previously been observed in other experiments such as the Rotating Wall Machine at the University of Wisconsin–Madison,\cite{paz-soldan2011} but can be challenging to measure as the contribution is typically much smaller than the background field.} It can be explained by considering the magnetic field line geometry in the system. As the current increases, the larger azimuthal field causes magnetic field lines to become more tightly wound. Magnetized electrons, flowing primarily along these field lines as they carry the rope current, follow this solenoidal path, contributing an additional magnetic field component which aligns with the background field, thus leading to a paramagnetic enhancement of the magnetic field at the center of each flux rope.

We can construct a simple analytical estimate of these two effects from our simulation's initial conditions. From the initial distribution functions defined by Eqs. \ref{eq:F0} and \ref{eq:G}, we find the perpendicular pressure profile for one rope,
\begin{align}
    &P_{e,\perp}(\vec{x}) = T_{e,r}n_{r,0}\exp(-r^2/r_s^2) + T_{e,b}n_{b,0},\\
    &P_{i,\perp}(\vec{x}) = T_{i,r}n_{r,0}\exp(-r^2/r_s^2) + T_{i,b}n_{b,0}.
\end{align}
The diamagnetic current is then found by
\begin{equation}
    \vec{J}_{\sigma,dia}(\vec{x}) = -\del \times\left(\frac{ P_{\sigma,\perp}\vec{B}}{B^2}\right).
\end{equation}
This current has both an azimuthal and axial component, but only the azimuthal component contributes to the diamagnetic magnetic field. We compute this contribution from Ampere's law as
\begin{align}
    \vec{\delta B}_{dia}(r=0) &= \mu_0 \int_0^\infty dr [\hat{\phi}\cdot\vec{J}_{\sigma,dia}(r)]\hat{z}\nonumber\\
    &= -\frac{\mu_0 n_r}{B_0}(T_{e,r} + T_{i,r}) \hat{z}.\label{eq:diamagnetic}
\end{align}

We may similarly estimate the paramagnetic contribution. As the magnetized electrons carrying the rope current follow field lines, they adopt an azimuthal component due to the spiraling of the field lines, i.e. $\vec{J}'=-en_rV_{0_e}\hat{b}$. Taking the azimuthal component and integrating as in Eq. \ref{eq:diamagnetic}, we find, for a single rope carrying total current $I$,
\begin{equation} \label{eq:paramagnetic}
    \vec{\delta B}_{para}(r=0) = \frac{\mu_0^2 \log{2}}{4B_0 \pi^2 r_s^2} I^2 \hat{z},
\end{equation}
giving a total central paramagnetic/diamagnetic contribution of
\begin{equation}\label{eq:paradia}
    \delta B_{max} = -\frac{\mu_0 n_r}{B_0}(T_{e,r} + T_{i,r}) + \frac{\mu_0^2 \log{2}}{4B_0 \pi^2 r_s^2} I^2.
\end{equation}

\begin{figure}[b]
    \centering
    \includegraphics[width=\linewidth]{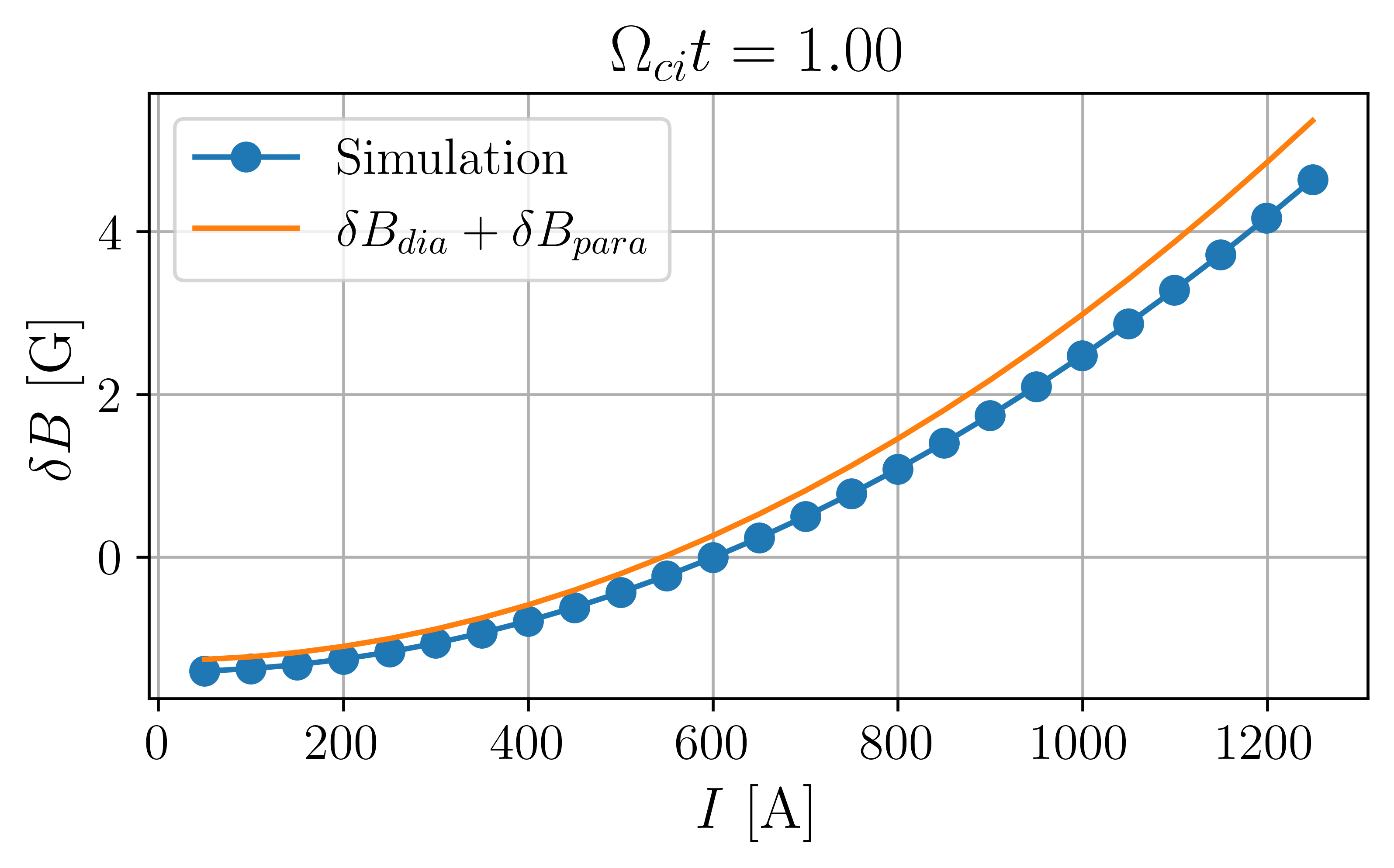}
    \caption{Azimuthal current-induced magnetic perturbation in the $\hat{z}$ direction for a single flux rope after initial electron-scale relaxation of the system as a function of the total current, from 2D simulations (blue) and Eq. \ref{eq:paradia} (orange). The transition from diamagnetism at low current to paramagnetism at high current occurs when $\delta B$ switches signs at $I\approx\qty{600}{\ampere}$ in simulations compared to $I\approx\qty{550}{\ampere}$ from Eq. \ref{eq:paradia}.}
    \label{fig:current_scan}
\end{figure}

To verify this model, we performed a series of two-dimensional, single rope simulations in the $x$-$y$ plane at 25 different values of $I$ ranging from $\qty{100}{\ampere}$ to $\qty{1250}{\ampere}$. Fig. \ref{fig:current_scan} shows the central magnetic field perturbation after initial relaxation occurs in these simulations and compares them with the predicted perturbation from Eq. \ref{eq:paradia} based on the simulation initial conditions, showing strong agreement. The transition from diamagnetism at low current to paramagnetism at high current occurs when $\delta B$ switches signs, at a total discharge current of $I\approx\qty{600}{\ampere}$ in simulations, compared to $I\approx\qty{550}{\ampere}$ from Eq. \ref{eq:paradia}.

\subsection{Ohm's Law Balance}
    \begin{figure*}[ht]
        \centering
        \includegraphics[width=\textwidth]{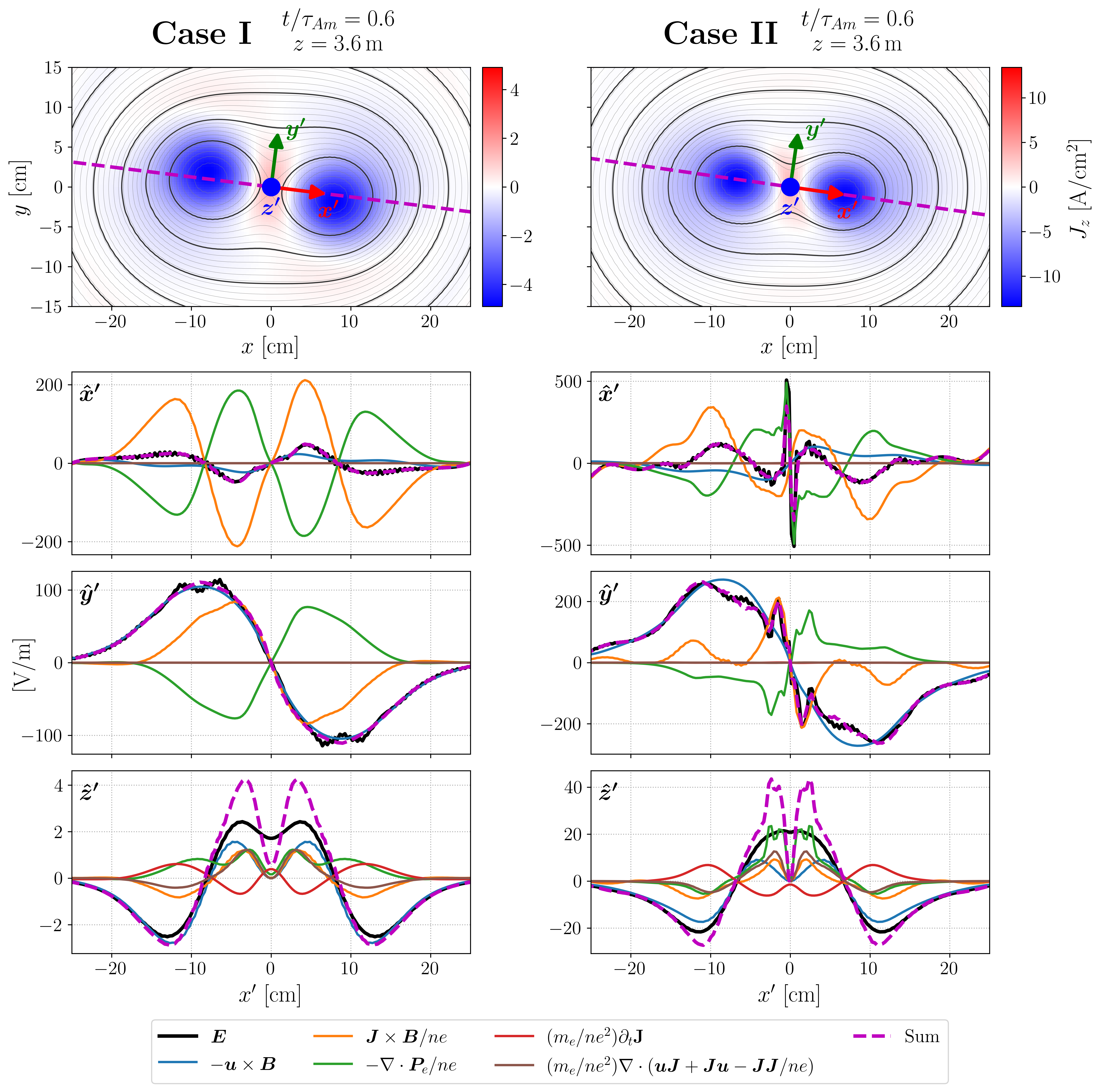}
        \caption{Comparison of electric field support in case I and case II. The top row shows out-of-plane current $J_z$ in an $x$-$y$ plane slice at $z=\qty{3.6}{\meter}$ at a normalized time of $t/\tau_{Am}=0.6$. The overlayed contours indicate surfaces of the flux function $\psi$. Term-by-term components of line-outs taken along the indicated magenta dashed line are shown for $x'$, $y'$, and $z'$ directions in the next three rows, respectively, as they vary along the $x'$ coordinate direction. We indicate the axes of this rotated coordinate system with arrows on the top row plots.}
        \label{fig:ohm_cartesian}
    \end{figure*}
The merging of two paramagnetic ropes at higher current in case II shows qualitative differences from that of diamagnetic ropes. In addition to the differences in timescale, the final helical twist that develops in the ropes is somewhat larger. Figs. \ref{fig:2d_cuts}II.a and \ref{fig:2d_cuts}II.b show the development of larger quadrupolar density structure, and Fig. \ref{fig:2d_cuts}II.c shows the paramagnetic current flows which develop as the ropes interact at early time. The current density in the flux ropes decays somewhat over the course of the simulation, which is why the ropes become diamagnetic at late time (\ref{fig:2d_cuts}II.f). 

With these differences in mind, it is worthwhile to investigate whether the merging which occurs in the two cases is quantitatively similar. To understand the physics governing the magnetic reconnection which occurs as the ropes merge, we consider the terms supporting the electric field from the generalized Ohm's law. Figure \ref{fig:ohm_cartesian} shows the contributions of each term in Eq. \ref{eq:ohms_law} to the total electric field along a cut taken between the points of maximum magnetic flux in each rope in an $x$-$y$ plane at $z=\qty{3.6}{\meter}$. The top row shows out-of-plane current density $J_z$ with contours of the 2D flux function $\psi$. The direction of the linear cut is indicated by the magenta dashed line. Along this line, we take the vector values of the terms in Eq. \ref{eq:ohms_law} and recast them in the primed coordinate system indicated by the arrows in the top row of plots, defined with $x'$ aligned with the cut direction. In the subsequent rows, we plot the components in the $\hat{x}'$, $\hat{y}'$, and $\hat{z}'$ directions as they vary along $x'$.

This analysis shows that the electric field is primarily supported by the inductive term in most of the domain, $\vec{E}\approx-\vec{u}\times\vec{B}$. This result is expected with the largest scales in the system being near or above the ion inertial length $d_i\approx\qty{0.3}{\meter}$. Near the quasi-separatrix layer, where reconnection is expected to be important, we see that the Hall and pressure terms become important to the support of $E_{z'}$ as the $\hat{x}'$ and $\hat{y}'$ components of the magnetic field vanish. While the additional inertial terms are nonzero near this region, they are nearly equal and opposite in magnitude, so are not important to the electric field support. 

The trace for $E_{z'}$ shows relatively large deviations from the sum of the supporting terms in Eq. \ref{eq:ohms_law} near the QSL. This deviation highlights an important limitation of the lowest-order PKPM model: true field-line breaking is facilitated by particles whose gyro-orbits cause them to pass through the separatrix and exhibit motion that deviates from the typical gyrotropic motion. While the electrons remain well-magnetized through most of the layer, there is nevertheless a point between the ropes where $B_x = B_y = 0$, leading to small deviations from gyrotropic motion which cannot be captured by the PKPM model. As previously mentioned in the discussion of initial conditions, we address this deficiency with a hyper-diffusive term which allows the electric field to be properly supported without modifying the kinetic response.\cite{juno2025} We have checked that the quantitative disagreement arises due to this hyperdiffusion parameter supporting the electric field at small scales but for clarity of comparison to the physical terms we elect to ignore the hyperdiffusion term in our Ohm’s Law analysis.

Here, we observe an apparent structural difference between the paramagnetic and diamagnetic ropes as reconnection occurs: In the diamagnetic case I, in the outer region away from the current sheet ($|x'|\gtrsim\qty8{\centi\meter}$), the pressure gradient term $-\del\cdot\vec{P}_e/ne$ provides a positive contribution and the Hall term $\vec{J}\times\vec{B}/ne$ provides a negative contribution to the total $E_{z'}$. These terms generally balance, leading to the electric field being entirely supported by the inductive term. However, in the paramagnetic case II, the sign of the pressure gradient term flips. Since the pressure gradient and Hall terms no longer balance, they provide an additional contribution to $E_{z'}$, leading to a larger electric field in the outer region than would be expected from the inductive term alone.

\begin{figure}[t]
    \centering
    \includegraphics[width=\columnwidth]{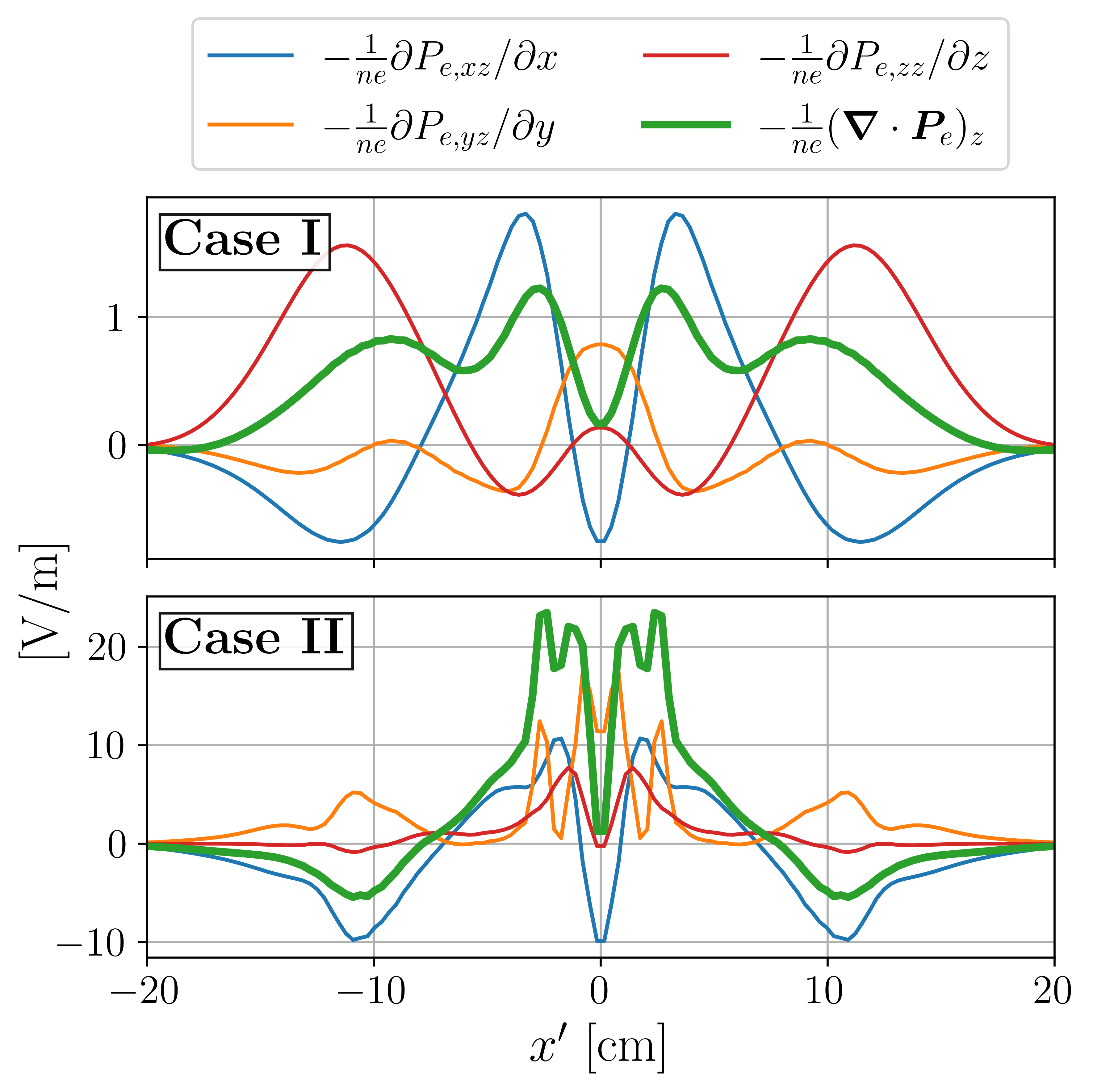}
    \caption{The $\hat{z}'$ component of the pressure gradient term of the electric field (green) shown in the cuts in Fig. \ref{fig:ohm_cartesian}, separated further into its individual pressure tensor component terms.}
    \label{fig:gradP_z}
\end{figure}

In the inner region, the Hall and pressure gradient terms are both positive in both cases. However, decomposing the pressure gradient term further shows how the cartesian components of the pressure tensor contribute differently. In Fig. \ref{fig:gradP_z}, we separate the term for the $\hat{z}'$ direction shown in Fig. \ref{fig:ohm_cartesian} into its individual cartesian tensor components. We see here that the largest difference between the two cases comes from the term $-(1/ne) \partial P_{e,zz}/\partial z$, i.e., a pressure gradient along the primary flow and the direction of the guide field. This term is relatively large in the outer region for case I, while it is much smaller than the other terms in case II. Moreover, the sign of this term flips entirely between the two cases in the inner region near the reconnection layer.

\subsection{Field-aligned Analysis and Reconnection Rate} \label{sec:recon_rate}
\begin{figure*}
    \centering
    \includegraphics[width=\textwidth]{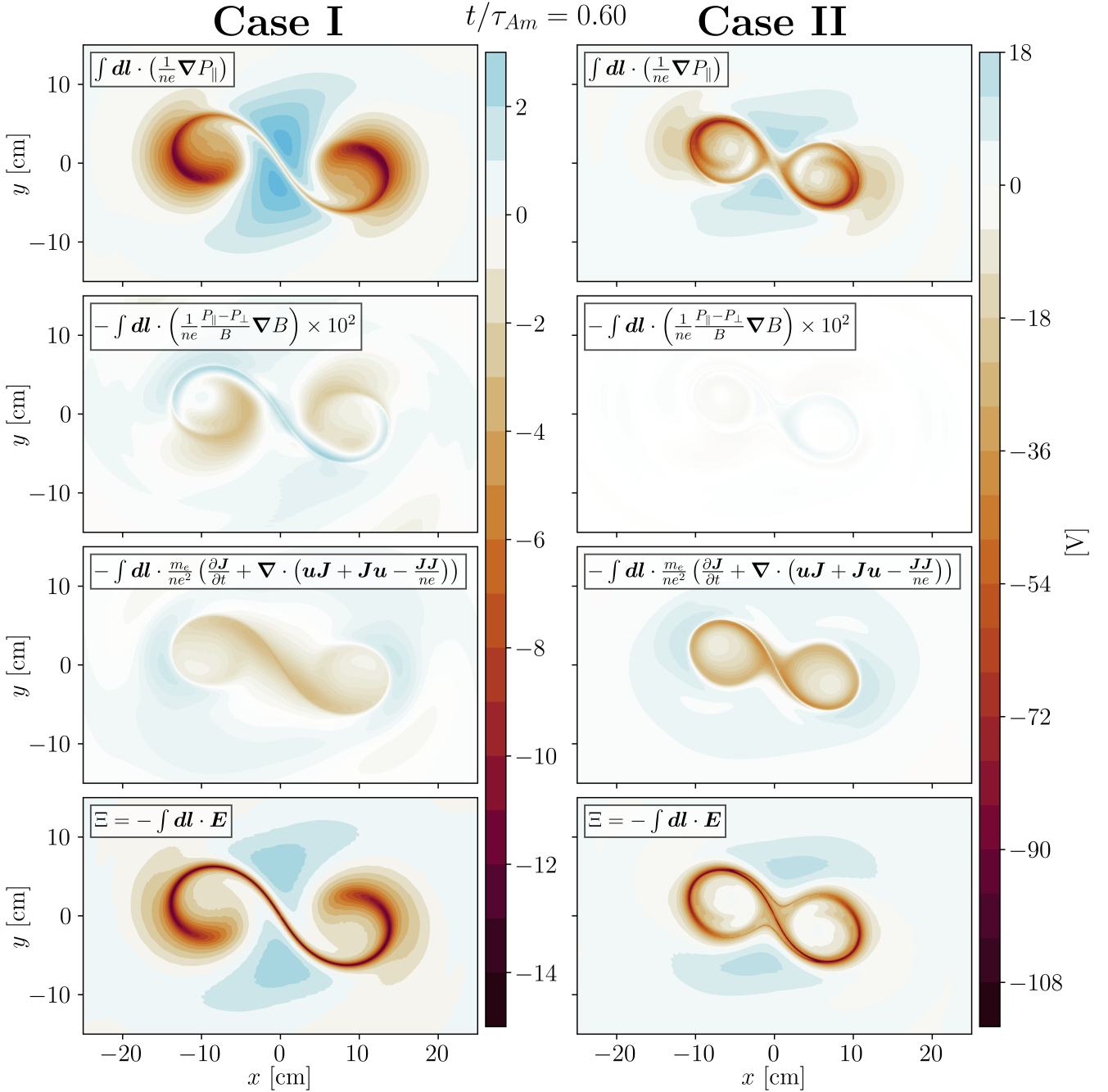}
    \caption{Integrated terms at $t/\tau_{Am} = 0.60$ in Eq. \ref{eq:quasipotential_ohm} for case I (left) and case II (right), in units of volts. Note that the pressure anisotropy term (second row) is very small; the term is multiplied by 100 here to make it visible using the same colormap scale.}
    \label{fig:int_ohm}
\end{figure*}
The structural differences noted in the previous section and shown in Figs. \ref{fig:ohm_cartesian} and \ref{fig:gradP_z} are a consequence of the Cartesian geometry used in the analysis. As the magnetic field vector tilts at a larger angle, particularly in case II, integrating along the reconnecting field lines captures essential contributions from the in-plane $\hat{x}$ and $\hat{y}$ components of the pressure tensor. Considering only the $\hat{z}$ component in a local coordinate frame, as is common in guide field reconnection analysis, ignores these contributions that are introduced by 3D magnetic geometry. 

The field-aligned quasi-potential diagnostic accounts for the complex magnetic geometry in these simulations and reveals similar underlying reconnection physics despite qualitative structural differences between case I and case II. Figure \ref{fig:int_ohm} demonstrates the structure of the quasi-potential $\Xi$ and its constituent terms per Eq. \ref{eq:quasipotential_ohm} at $t/\tau_{Am} = 0.60$, near the peak reconnection rate. We find that the structure of the layer highlighted by the quasi-potential is identical to that of the QSL.  The relatively reduced guide field in case II means that an individual field line may approach a region of reconnection multiple times across the domain. This interaction is reflected in the quasi-potential and QSL calculations, showing the structure that fully describes the two ropes as field lines converge and diverge multiple times across the domain. This behavior is in contrast to case I, in which the stronger relative guide field results in field lines generally completing less than one "orbit" before reaching the end of the domain, limiting the spatial extent of the structure in $Q$ and $\Xi$. 

The reconnection quasi-potential here reveals that the reconnection in our simulations is supported identically across the entire QSL in both case I and case II by the parallel pressure gradient, with small or negligible contributions from the anisotropy and inertial terms. The largest disagreement in the innermost part of the reconnection layer is due to the contribution from the hyperdiffusion term as previously discussed. 

\begin{figure}
    \centering
    \includegraphics[width=\columnwidth]{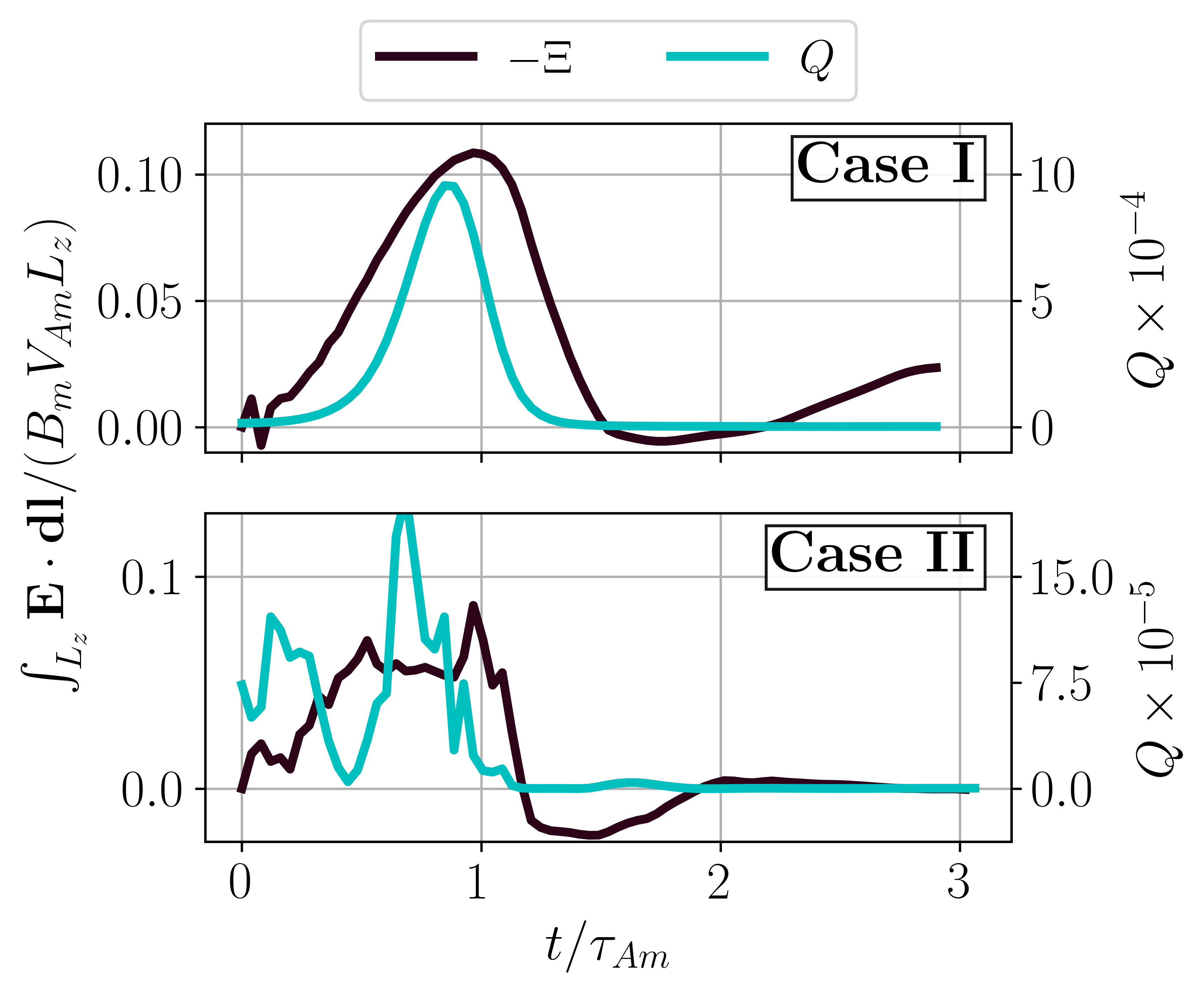}
    \caption{Normalized quasi-potential (black) and squashing factor (cyan) near $x=y=0$ as a function of time in case I (top) and case II (bottom).}
    \label{fig:recon_rate}
\end{figure}

Figure \ref{fig:recon_rate} shows the time-dependence of the quasi-potential $\Xi$, normalized to $B_m V_{Am}L_z$, and the squashing factor $Q$ calculated via Eqs. \ref{eq:squashing_factor} and \ref{eq:quasipotential} near the center of the simulation domain $x=y=\qty{0}{\meter}$. In both simulations, the reconnection rate, quantified by the normalized quasi-potential, reaches a peak value of $0.1$ at similar normalized times. The squashing factor also peaks at a similar time, demonstrating that both diagnostics are robust in quantifying the reconnection rate despite the differing magnetic geometry affecting the results of local coordinate analysis.

\section{Conclusions} \label{sec:conclusions}
We performed 3D simulations using the parallel-kinetic-perpendicular-moment (PKPM) model to study the physics of reconnecting flux ropes at experimental parameters. We identified a current dependence on the initial relaxation of individual flux ropes, which causes the ropes to transition between diamagnetic at low current densities and paramagnetic at high current densities. These two cases show different large-scale structure, with the higher-current paramagnetic case leading to a more pronounced helical structure of the two ropes at late time. In the Cartesian frame, the pressure tensor term supporting the electric field appears to play a different role between the two regimes, with the term flipping signs entirely on the outer side of the merging flux ropes and different pressure tensor components becoming important in the reconnection layer. We find, however, that these apparent differences are a consequence of the magnetic geometry of the system. Field-aligned 3D diagnostics—specifically the quasi-potential $\Xi$ and squashing factor $Q$—shows that the underlying kinetic-scale physics of reconnection are the same in these two systems, and that the parallel pressure gradient provides the primary reconnection support in the QSL. Examining the time-dependence of these diagnostics shows that they are robust measures of reconnection rate. 

Our findings show the importance of accounting for the magnetic morphology when studying magnetic reconnection in 3D systems. Considering only local physics may lead to incorrect physical interpretations, but global diagnostics provide a robust analysis of the magnetic reconnection rate despite the complexities involved in 3D systems. The PKPM model provides a powerful new approach to studying these magnetized systems by separating the kinetic parallel physics of reconnection from the more fluid-like perpendicular motion. 

We emphasize that the simulations presented here were performed at \textit{experimental} parameters. In contrast to other simulation work which typically attempts to preserve a limited number of dimensionless quantities in order to connect to experimental systems, we here directly simulate the length scales, time scales, and plasma parameters relevant to laboratory experiments, with the exception of the parameters modified for computational tractability as discussed in Sec. \ref{sec:init_conds}. In particular, we highlight that the Debye length in our simulations, $\lambda_D\approx\qty{1.5e-4}{\meter}$, is highly under-resolved by the spatial grid resolution, $\Delta x=\Delta y\approx\qty{6.3e-3}{\meter}$ and $\Delta z \approx\qty{7e-2}{\meter}$. Matching the parameters used in this study would be impossible in a PIC simulation without massively increased resolution (and therefore computational expense). Nevertheless, these systems remain fundamentally kinetic, and moving to cheaper fluid plasma models omits important physics. The PKPM system makes these simulations computationally affordable, \textcolor{black}{allowing simulations to span scales large enough to demonstrate MHD-scale plasma dynamics while still including the microphysics necessary to describe magnetic reconnection and other kinetic-scale phenomena.}

The current-dependent transition from diamagnetism to paramagnetism we observe and explain in the simulations presented here carries some relevance to other flux rope merging experiments in the field. Electron-scale merging experiments at the PHASMA device at West Virginia University observed diamagnetic flux ropes \cite{shi2021} which rotate and merge faster than the ideal kink mode growth \cite{shi2022}, as in our simulations. The experiments further observed that the flux rope rotation direction would reverse as the discharge current ramped up. This time-dependent discharge current would correspond to the transition between diamagnetism and paramagnetism we derived in Eq. \ref{eq:paradia}, suggesting that it may play a role in the observed rotation, and we plan on future work using our model to explain these observations. 

\textcolor{black}{In its current form, the PKPM model includes self-collisions, but the collision frequency is not velocity-dependent, and inter-species collisions are not implemented. It is likely that including inter-species collisions would have important effects in our simulations.} In particular, LAPD experiments observe flux rope and current sheet heating \cite{gekelman2012} in which plasma resistivity likely plays an essential role. Future work to expand the collision operator modeling capabilities in PKPM will help clarify details on the role resistivity plays in these systems. 

\begin{acknowledgments}
J.J. gratefully acknowledges the \texttt{Gkeyll}~team for fruitful discussions, especially A. Hakim. This work was supported by the U.S. Department of Energy under contract number DE-AC02-09CH11466 via LDRD grants. J.J. and  J.M.T., and the development of \texttt{Gkeyll}, were partly funded by the NSF-CSSI program, Award No. 2209471. J.M.T. was also supported by NSF award AGS-2401110. The simulations and analysis presented in this article were performed in part on computational resources managed and supported by Princeton Research Computing, a consortium of groups including the Princeton Institute for Computational Science and Engineering (PICSciE) and the Office of Information Technology’s High Performance Computing Center and Visualization Laboratory at Princeton University. This research is part of the Frontera computing project at the Texas Advanced Computing Center. Frontera is made possible by National Science Foundation (NSF) Award OAC-1818253.
\end{acknowledgments}

\section*{Data Availability Statement}
    The input files for the \texttt{Gkeyll} simulations presented here are available in the following GitHub repository, \url{https://github.com/ammarhakim/gkyl-paper-inp}. All plots in this paper utilize the \texttt{postgkyl} package \url{https://github.com/ammarhakim/postgkyl}.

\section*{Author Declarations}
\subsection*{Conflict of Interest}
The authors have no conflicts to disclose.

\subsection*{Author Contributions}
\textbf{Joshua Pawlak}: Conceptualization (equal); Data curation (lead); Formal analysis (lead); Formal analysis (lead); Investigation (equal); Methodology (equal); Resources (equal); Software (equal); Validation (equal); Visualization (lead); Writing - original draft (lead); Writing - review \& editing (equal). \textbf{James Juno}: Conceptualization (equal); Data curation (equal); Formal analysis (equal); Funding acquisition (equal); Investigation (equal); Methodology (equal); Project administration (equal); Resources (equal); Software (lead); Supervision (equal); Validation (equal); Visualization (equal); Writing - original draft (equal); Writing - review \& editing (equal). \textbf{Jason M. TenBarge}: Conceptualization (equal); Data curation (equal); Formal analysis (equal); Funding acquisition (equal); Investigation (equal); Methodology (equal); Project administration (equal); Resources (equal); Software (equal); Supervision (equal); Validation (equal); Visualization (equal); Writing - original draft (equal); Writing - review \& editing (equal).
\appendix
\section{Details on Initial Conditions} \label{sec:init_details}

Equations \ref{eq:F0} and \ref{eq:G} are sufficient to completely close the initial condition system, provided that $\vec{\del}\times\vec{B} = \mu_0 \vec{J}$. However, while this system is correct, for completeness and to illustrate the subtlety in the way the distribution function is defined in PKPM, we use this appendix to demonstrate the entire initial condition system exactly as it is implemented numerically in our simulations. The complete list of simulation parameters is supplied in Table~\ref{tab:params}.

The important note is that the distribution functions for the counter-streaming background and rope populations given in Eqs. \ref{eq:F0} and \ref{eq:G} are written for the \textit{lab-frame} parallel velocity $v_\parallel$. As it is introduced in Ref.~\onlinecite{juno2025}, the distribution functions used in the PKPM system are defined in the frame moving with the bulk fluid velocity for each species. To correctly transform into this frame, we write $\vec{v} = \vec{u} + \vec{v}'$, where $\vec{u}$ is the species fluid velocity. Since the ions are stationary, their distribution function form remains unchanged. For electrons: 
\begin{align}
    \begin{split}
    \vec{u}_e(\vec{x}) &= \frac{\int dv_\parallel v_\parallel \left[ F_{0_{e,b}}(\vec{x},v_\parallel') + F_{0_{e,r}}(\vec{x},v_\parallel') \right]}{\int dv_\parallel \left[ F_{0_{e,b}}(\vec{x},v_\parallel') + F_{0_{e,r}}(\vec{x},v_\parallel') \right]} \\
    &=\frac{n_r(\vec{x}) \vec{V}_{0_e}}{n_b(\vec{x}) + n_r(\vec{x})}.
    \end{split}
\end{align}
Then, Eqs. \ref{eq:background} and \ref{eq:rope} become 
\begin{align}
    &F_{0_{e,b}}(\vec{x},v_\parallel') = n_b(\vec{x})\sqrt{\frac{m_e}{2\pi T_{e,b}}}\exp\left[-\frac{m_e (v_\parallel' + \hat{b}\cdot\vec{u}_e)^2}{2T_{e,b}}\right], \\
    \begin{split}
    &F_{0_{e,r}}(\vec{x},v_\parallel') = n_r(\vec{x})\sqrt{\frac{m_e}{2\pi T_{e,r}}}
    \\&\phantom{F_{0_{e,r}}(\vec{x},v_\parallel') = }\times\exp\left[-\frac{m_e}{2T_{e,r}}(v_\parallel' + \hat{b}\cdot\vec{u}_e - \vec{V}_{0_e}\cdot\hat{b})^2\right].
    \end{split}
\end{align} 

Using Eq. \ref{eq:magfield}, which gives the azimuthal magnetic field about the center of a single rope, we may calculate $\hat{b}=\vec{B}/|\vec{B}|$ for a superposition of two ropes at locations $\vec{x}_1$ and $\vec{x}_2$. In this case, the magnetic field vector in Cartesian coordinates becomes

\begin{align}
    \begin{split}
    &\vec{B}(\vec{x}) = -[B_1(\vec{x})\sin{\phi_1} + B_2(\vec{x})\sin{\phi_2}]\hspace{2pt}\hat{x}\\
    &\phantom{\vec{B}(\vec{x}) =} + [B_1(\vec{x})\cos{\phi_1} + B_2(\vec{x})\cos{\phi_2}]\hspace{2pt}\hat{y}+ B_0\hat{z},
    \end{split}
\end{align}
where $\phi_1$, $\phi_2$ are the angles between $\hat{x}$ and ($\vec{x} - \vec{x}_{1}$) and ($\vec{x} - \vec{x}_{2}$), respectively, and
\begin{align}
    &B_1(\vec{x}) = \left[ 1 - \exp\left(-\frac{|\vec{x} - \vec{x}_{1}|^2}{r_s^2}\right) \right]\frac{\mu_0 I_{1}}{2\pi |\vec{x} - \vec{x}_{1}|}\\
    &B_2(\vec{x}) = \left[ 1 - \exp\left(-\frac{|\vec{x} - \vec{x}_{2}|^2}{r_s^2}\right) \right]\frac{\mu_0 I_{2}}{2\pi |\vec{x} - \vec{x}_{2}|}.
\end{align}

Note that with this frame shift, we ensure that the first velocity moment $\int dv_\parallel' v_\parallel' F_{0_e}(v_\parallel') = 0$, demonstrating that the fluid motion in PKPM is entirely contained within the fluid equations solved separately from the reduced kinetic Eqs. \ref{eq:pkpmF} and \ref{eq:pkpmG}. 

We also add an additional magnetic field perturbation to the initial conditions in the $\hat{z}$ direction to balance the outward pressure force due to the top-hat density profile of the background in Eq. \ref{eq:background}. This field is found simply by balancing the magnetic and plasma pressures in the inner and outer regions. In the limit $\delta B_z\ll B_0$, this becomes
\begin{equation}
    \delta B_z(r) = \frac{\mu_0 n_{b,0}}{2 B_0}(T_{e,b} + T_{i,b})\left[ 1 + \tanh\left( \frac{r - L_r}{\rho_s} \right) \right],
\end{equation}
which gives a slightly larger field in the outer region of the simulation. This perturbation prevents the background plasma column from expanding in the initial relaxation, which may otherwise affect the flux rope dynamics.

\begin{table}[t]
\caption{\label{tab:params}Full list of simulation initial condition parameters.}
\begin{tabular}{c|c}
    Parameter & Value \\
    \hline
    $n_{b,0}$ &  \qty{2e18}{\meter^{-3}} \\ 
    $n_{r,0}$ & \qty{1e18}{\meter^{-3}} \\
    $T_{e,b}$ & \qty{5}{\electronvolt}\\
    $T_{e,r}$ & \qty{17.3}{\electronvolt}\\
    $T_{i,b}$ & \qty{1}{\electronvolt}\\
    $T_{i,r}$ & \qty{3.4}{\electronvolt}\\
    $m_i$ & $4 m_p$\\
    $m_i/m_e$ & 400\\
    $I_0$ & \qty{-750}{\ampere} / \qty{-2000}{\ampere}\\
    $B_0$ & \qty{330}{\gauss}\\
    $r_s$ & \qty{4.45}{\centi\meter}\\
    $L_r$ & \qty{30}{\centi\meter}\\
    $|\vec{x_1} - \vec{x_2}|$ & \qty{22.1}{\centi\meter}\\
    $a$ & 0.1\\
    $\epsilon_{sim}$ & $1000\epsilon_0$ \\
    $\nu_{hyp}$ & $10^{-3}$\\
\end{tabular}
\end{table}

{\color{black}
\section{Flux Rope Kink Modes} \label{sec:kink_modes}

The LAPD experiments on which we base our simulations note that the flux ropes are kink-unstable and report a characteristic oscillation frequency consistent with theoretical analysis of kink modes in line-tied flux ropes. \cite{gekelman2016a,ryutov2006} This theory predicts that the flux ropes in our simulations are also kink-unstable, with growth rate and real frequency of the instability given by \cite{ryutov2006}
\begin{equation}
    \tan\left( L_z k_0\sqrt{\frac{1}{4} + \frac{\omega^2}{2{v_A}^2}} \right) = -2i\sqrt{\frac{1}{4} + \frac{\omega^2}{2{v_A}^2}},
\end{equation}
where $k_0 = \frac{B_{\phi,max}}{r_s B_z}$, $L_z$ is the domain length, and $v_A$ is the standard definition of the Alfv\'en speed. Using the initial conditions for case I, we solve this equation numerically to find a solution $\omega=\omega_r + i\gamma$, where $\omega_r\approx-\qty{3.4e4}{\per\second}$ and $\gamma\approx\qty{6.2e4}{\per\second}$, which is comparable to the solution found for the LAPD experiment. \cite{gekelman2016a}

Here, $\gamma$ corresponds to a linear $e$-folding growth time of $\tau\approx\qty{16}{\micro\second} \approx 1.1\tau_{Am}$. Fig. \ref{fig:recon_rate} shows that the time of peak reconnection rate in case I, as measured by the quasi-potential $\Xi$, is $t\approx1.0\tau_{Am}$, but also that reconnection proceeds almost immediately in our simulation. The physical displacement of the plasma as merging proceeds within this time is also large. This suggests that, while an ideal kink instability may be present in the system, it does not have sufficient growth time to be the dominant driver of the reconnection in our simulation. Instead, the large-scale $\vec{J}\times \vec{B}$ force exerted on each flux rope by the other due to their parallel currents is sufficient to explain the merging process.

Our simulation is initialized with two pristine flux ropes which are individually unperturbed at $t=0$ and span the full length of the domain, which gives a strong mutual attractive force. Such ideal initial conditions may be unfeasible to realize in the lab; various perturbations to the plasma, particularly during the initial flux rope formation process, may facilitate the more rapid growth of a kink mode in a real system. This may further explain why such a mode is not observed in our simulations despite being readily observed in laboratory experiments.

In any case, the timescales for large-scale driving forces, whether a kink mode or mutual $\vec{J}\times\vec{B}$ force, are slower than the reconnection timescale. Our analysis of the quasi-potential and reconnection rate agrees with experimental findings,\cite{gekelman2016a} suggesting that rope merging and reconnection itself proceeds similarly irrespective of differences in the initial driving forces.

To demonstrate what the kink mode may look like at late time in our simulations, we perform an additional 3D PKPM simulation of a single flux rope (using the same initial conditions given in Sec. \ref{sec:model} for case II) with additional localized helical perturbations to the radial coordinate of the flux rope center of the form 
\begin{align}
    \begin{split}
    \vec{\xi}(\vec{r}) &= \xi_0\exp\left(-r^2/r_s^2 \right) \left(1-\exp\left(-z^2 \right)\right) \\
    &\times\left[\cos\left(\frac{2\pi k z}{L_z}  + \tilde{\phi}_k\right)\hat{x} + \sin\left(\frac{2\pi k z}{L_z}  + \tilde{\phi}_k\right)\hat{y}\right]
    \end{split}
\end{align}
summed for wavenumbers from $k=1$ to $k=3$, with $\xi_0 = 0.07r_s$ and a randomized phase shift $\tilde\phi_k$. Since we are interested in MHD modes, we hold the ion mass at the real value of helium and set the electron mass ratio as before at $m_i/m_e = 400$. We discretize a physical domain of $\qty{0.6}{\meter}\times\qty{0.6}{\meter}\times\qty{10}{\meter}$ (with the same $v_\parallel$ domain as before) on a $64\times64\times140\times16$ grid.

We show in Fig. \ref{fig:kink_mode} that, as expected, the flux rope at late time shows structure consistent with that of an $m=1$ ideal kink mode. We note, however, that the timescale over which this growth occurs is substantially slower than the timescales considered in the case I and II simulations we present in Sec. \ref{sec:results}. Despite the artificially large initial perturbation, the largest axial displacement from the initial center of the rope of \texttilde$\qty{2}{\centi\meter}$ is still much less than the displacement observed in the merging simulation at a similar time (\texttilde$\qty{15}{\centi\meter}$). Thus, while the physics necessary for a kink mode is present in our simulations, such modes should not be expected to grow large within the allocated simulation runtime presented here and are not necessary to explain the merging process we observe.


\begin{figure}
    \centering
    \color{black}
    \includegraphics[width=\columnwidth]{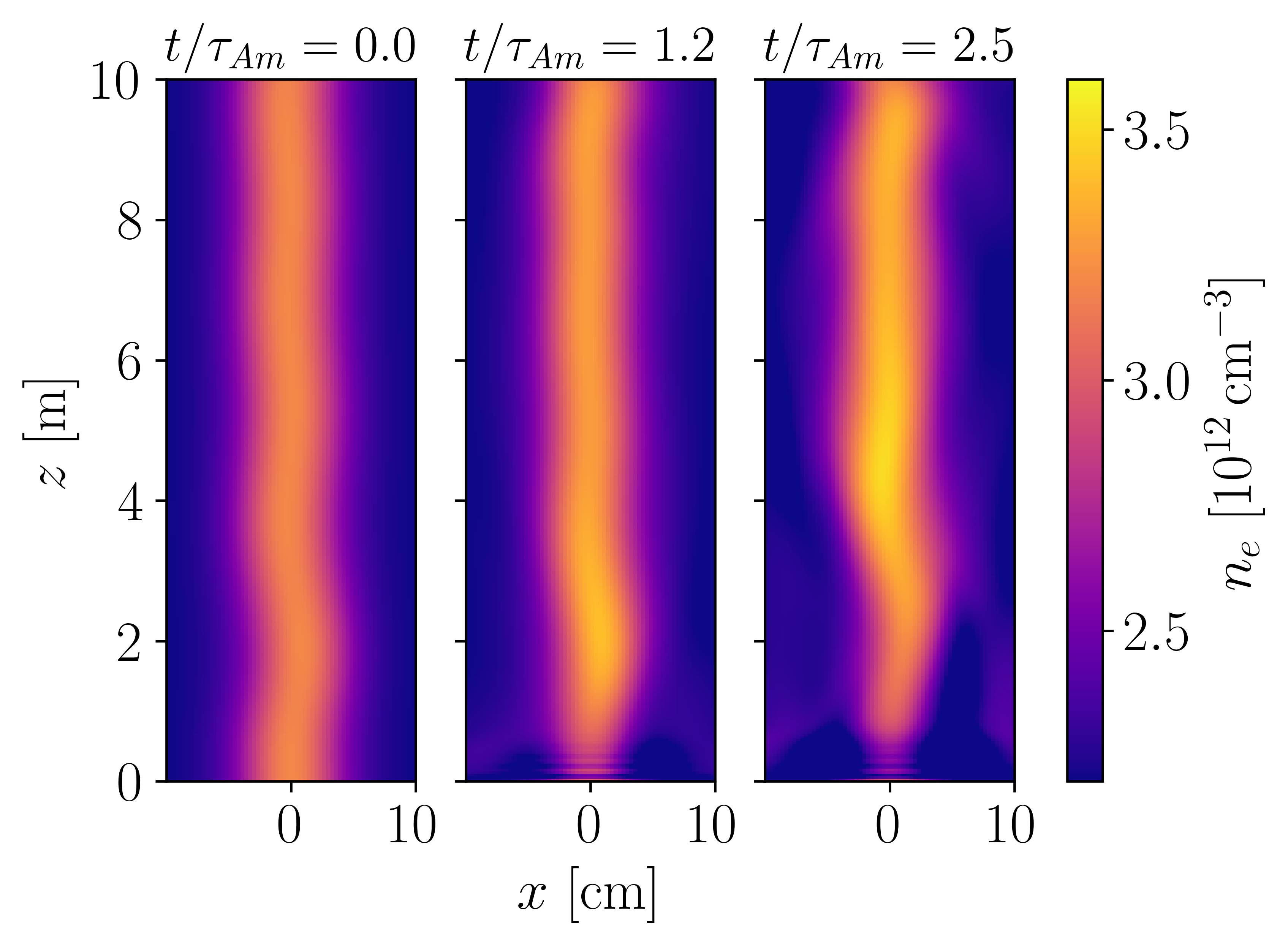}
    \caption{Plot of electron density along $x$-$z$ slices at $y=0$ for three time snapshots in the single kink-unstable flux rope simulation.}
    \label{fig:kink_mode}
\end{figure}

}

{\color{black}
\section{Notes on the Squashing Factor Calculation} \label{sec:squashing}

The numerical value of the squashing factor (Eq. \ref{eq:squashing_factor}) is highly sensitive to the details of the method used to compute it when considering discretized magnetic field measurements as in our simulations. We use this Appendix to illustrate the subtly in calculating and interpreting this quantity.

\subsection{Effects of Numerical Choices}
Computing Eq. \ref{eq:squashing_factor} requires computing numerical derivatives in the field line map $(x,y)\rightarrow(X,Y)$ for $(x,y)$ and $(X,Y)$ on the planes $z=z_0$ and $z=z_f$, respectively. As noted in Sec. \ref{sec:3d_diagnostics}, we do this by initializing a grid of seed points on the plane $z_0$ and tracing field lines originating at these locations to their final coordinates on the plate $z_f$ using the interpolated magnetic field values along those paths. Because field lines within regions of high $Q$ diverge rapidly from each other, the corresponding $(X,Y)$ coordinates are much more spread out than the seed point grid separation, placing a numerical limit on the maximum peak in $Q$ that may be computed at a given resolution. As noted in experimental studies on QSL structure in LAPD, in order to fully resolve the most strongly-peaked structure of $Q$, a substantially greater resolution must be used.\cite{gekelman2012}

\begin{figure}
    \centering
    \color{black}
    \includegraphics[width=\columnwidth]{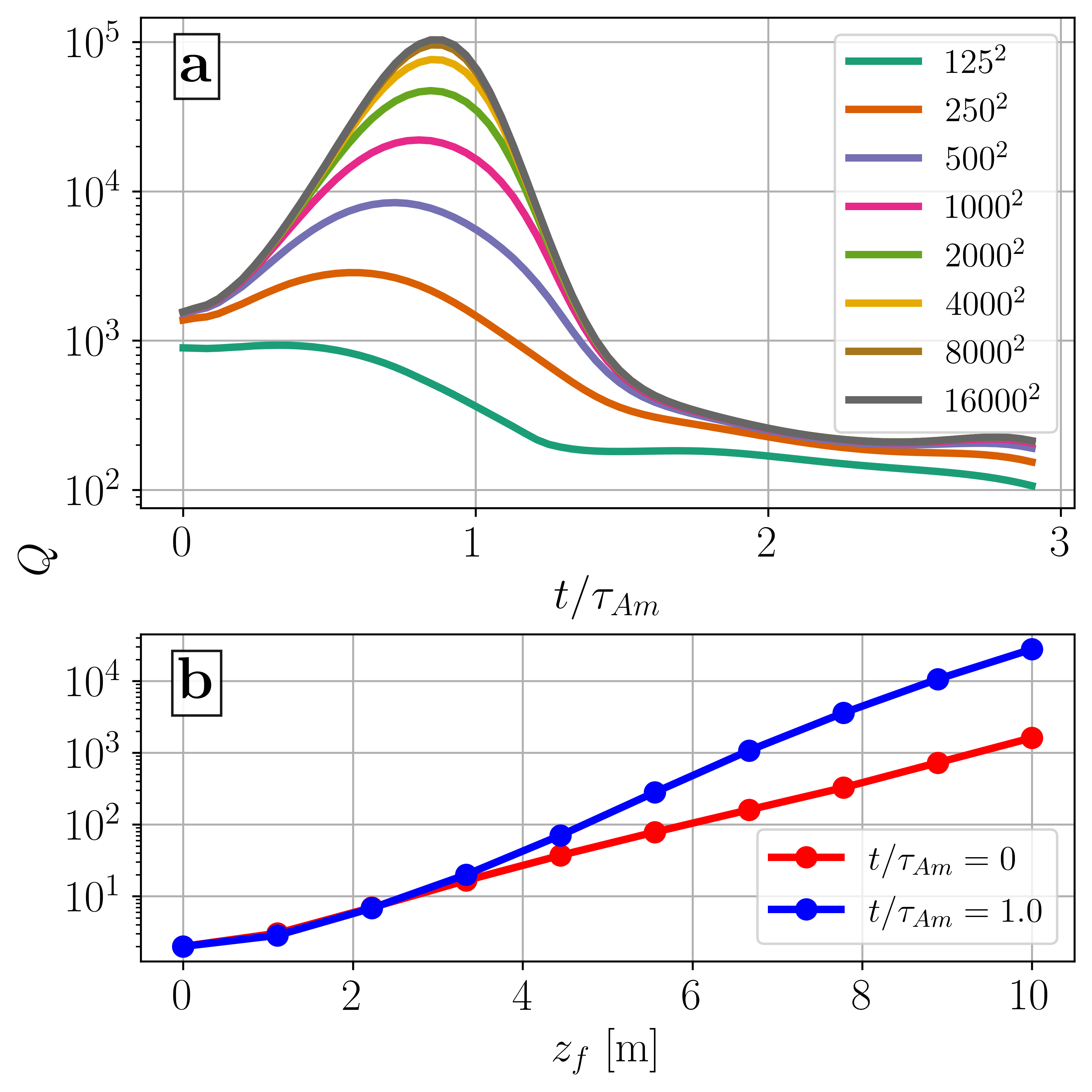}
    \caption{Numerically computed values of $Q$ for the case I simulation, showing dependence on (a) the indicated seed point resolution in $x$ and $y$ and (b) the choice of $z_f$ for a fixed $4000\times4000$ seed point resolution and $z_0 = \qty{0}{\meter}$ for the initial conditions and a time when $Q$ is large.}
    \label{fig:Q_res}
\end{figure}


In Fig. \ref{fig:Q_res}a, we show a plot of $Q$ versus time near the center of the domain calculated at several seed point resolutions as a function of time for case I. We show that the apparent maximum value of $Q$ can be as much as two orders of magnitude lower than the true maximum when calculated at sufficiently high resolution. Similarly, the computed value of $Q$ is dependent on the choices of $z_0$ and $z_f$. In our system, we consider the full extent of the domain, $z_0 = \qty{0}{\meter}$ and $z_f=\qty{10}{\meter}$. However, Fig. \ref{fig:Q_res}b shows that calculating $Q$ with a different choice of $z_f <\qty{10}{\meter}$ can reduce the numerically-computed value of $Q$ significantly, particularly at $t/\tau_{Am}=1.0$ which is when the reconnection rate is large. The result here is consistent with the understanding that field lines diverge exponentially where $Q$ is large.

Nonetheless, the qualitative structure of the layer is unaffected by resolution or choice of $z_f$, and the \textit{time-dependent} behavior of $Q$ agrees across all resolutions shown beyond $250\times250$ and is largest at the time of maximum reconnection rate (see Fig. \ref{fig:recon_rate}). Moreover, the structure of the quasi-potential $\Xi$, seen in Fig. \ref{fig:int_ohm}, agrees very well with the QSL structure even when defined with a relatively low value of $Q$. We conclude that the physics content of this diagnostic is robust to the numerical details of seed point resolution and $z$ domain extent. The absolute magnitude of the diagnostic is not important; its interpretation is contained in its general structure and relative behavior in time.

\subsection{Effects of Magnetic Geometry}
One may ask why a region of large $Q$ would be present at $t=0$, when no reconnection has yet occurred. This scenario is explained by considering that $Q$ is calculated purely from the magnetic geometry at a given time. While $Q$ tends to be large in reconnecting plasmas where field lines converge and diverge rapidly, other systems may have large $Q$ despite no reconnection occurring.

In fact, $Q$ can be in general quite large in any system which contains magnetic shear. As a relevant example, we show in Fig. \ref{fig:single_rope_q} the structure of $Q$ at $t=0$ for the single kink-unstable flux rope system described above in Appendix \ref{sec:kink_modes}. This calculation shows that, $Q$ can be relatively large (here, $Q_{max}\approx440$) even in a non-reconnecting plasma. This behavior is purely attributed to the shearing of magnetic fields between concentric flux surfaces in the rope. Caution must therefore be taken when taking the presence of (or lack of) a large $Q$ alone to characterize magnetic reconnection. The squashing factor is a measure of the magnetic configuration, not of reconnection itself.

\begin{figure}
    \centering
    \color{black}
    \includegraphics[width=\columnwidth]{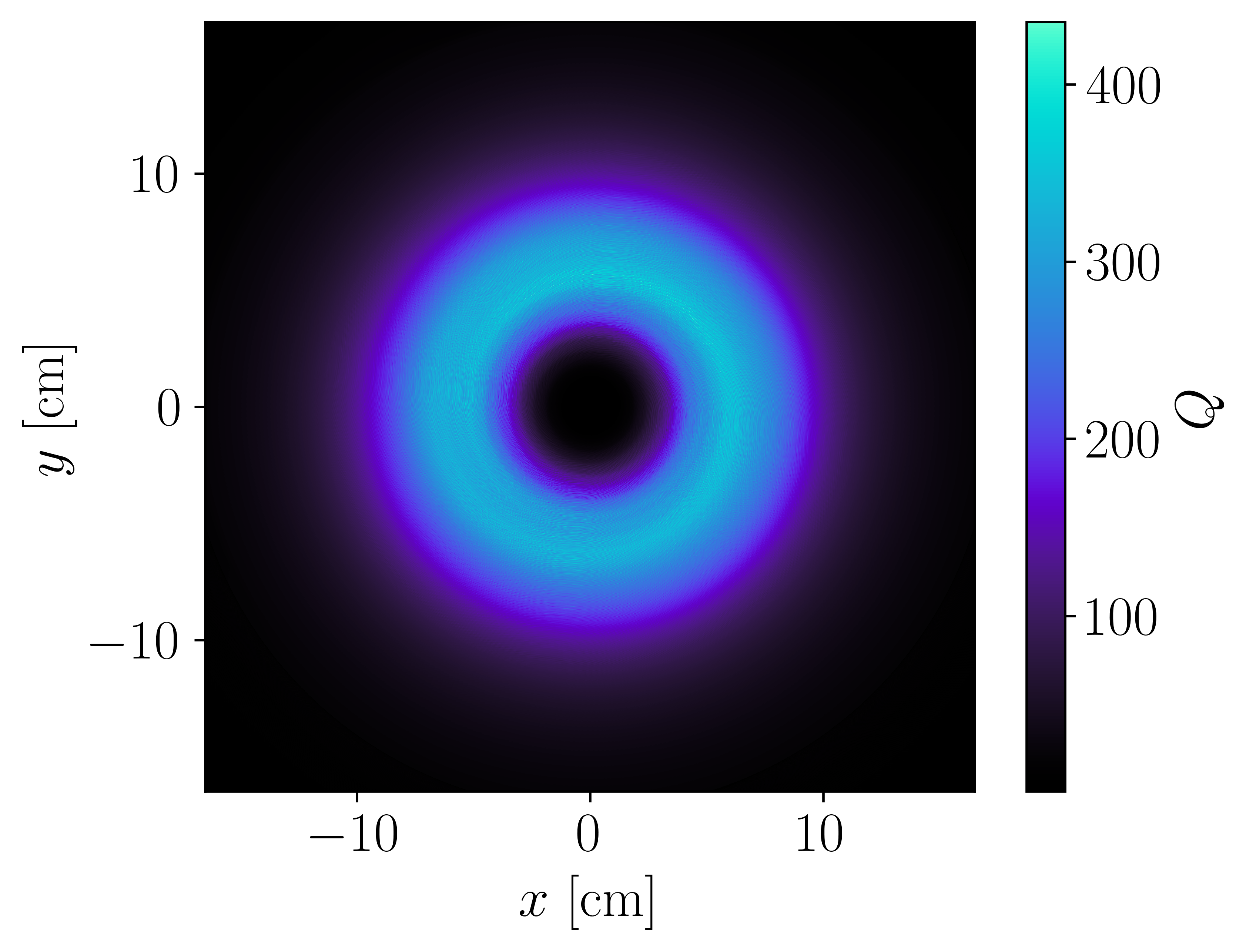}
    \caption{Squashing factor $Q$ computed on a $4000\times4000$ seed point grid for a single flux rope with a helical perturbation.}
    \label{fig:single_rope_q}
\end{figure}

}

\nocite{*}
\bibliography{citations}

\end{document}